\documentclass[useAMS,usenatbib]{mnras}
\usepackage{graphicx}
\usepackage{epsfig}
\usepackage{psfig}

\newcommand{\bfig}{\begin{figure*}}
\newcommand{\efig}{\end{figure*}}
\newcommand{\btab}{\begin{table*}}
\newcommand{\etab}{\end{table*}}
\newcommand{\bcen}{\begin{center}}
\newcommand{\ecen}{\end{center}}

   \title{Impact of photometric variability on age and mass determination of Young Stellar Objects: \\A case study on  Orion Nebula Cluster}

\author[Messina Sergio, et al.]{Sergio Messina$^{1}$\thanks{E-mail: sergio.messina@oact.inaf.it}, 
Padmakar Parihar$^{2}$\thanks{E-mail: psp@iiap.res.in},
  Elisa Distefano$^{1}$\thanks{E-mail: elisa.distefano@oact.inaf.it}\\
$^{1}$INAF-Catania Astrophysical Observatory, via S. Sofia 78, I-95123 Catania, Italy\\
$^{2}$Indian Institute of Astrophysics, Bangalore 560034, India\\}
\begin{document}

\date{Accepted ..... Received ......; in original form .....}
\pagerange{\pageref{firstpage}--\pageref{lastpage}} \pubyear{....}

\maketitle

\label{firstpage}

\begin{abstract}

In case of pre-main sequence objects,  the only way  to determine age
and  mass  is by  fitting  theoretical  isochrones on  color-magnitude
(alternatively  luminosity-temperature) diagrams.  Since  young stellar
objects exhibit  photometric variability over wide  range in magnitude
and  colors, the  age and  mass  determined by  fitting isochrones  is
expected to  be inaccurate, if not erroneous.  These \rm in  turn will badly  affect any study
carried out on age spread and process of star formation. Since we have
carried out very extensive photometric observations of the Orion Nebula Cluster (ONC), we decided
to use our multi-band data to explore the influence of variability in
determining mass  and age of cluster  members.  In this  study, we get
the amplitudes of  the photometric variability in V,  R, and I optical
bands of a sample of 346 ONC members and use it to investigate how the
variability affects the  inferred masses and ages and  if it alone can
take  account for the  age spread  among the  ONC members  reported by
earlier studies.  We  find that members that show  periodic and smooth
photometric   rotational  modulation  have   their  masses   and  ages
unaffected by variability.  On other  hand, we found that members with
periodic  but  very scattered  photometric  rotational modulation  and
members with irregular variability   have  their   masses   and  ages   significantly
affected. Moreover, using Hertzsprung-Russell (HR) diagrams we find that the observed I band
photometric  variability   can  take   account  of  only   a  fraction
($\sim$50\%)  of   the  inferred  age  spread,  whereas   the  V  band
photometric variability is large enough to mask any age spread.

\end{abstract}

\begin{keywords}
stars: activity - stars:  rotation - stars: pre-main-sequence - stars:
Hertzsprung-Russell and colour Ðmagnitude diagrams - stars: individual:
Orion Nebula Cluster

\end{keywords}

\section{Introduction}

Almost all Orion Nebula Cluster  (ONC) members are variable stars, the
amplitude of their optical  photometric variability ranging from a few
hundredths    up   to    more   than    1   magnitude    (see,   e.g.,
\citealt{Herbst02}). Generally, classical T Tauri stars (CTTS) exhibit
variability levels  larger than weak  line T Tauri stars  (WTTS) (see,
e.g.,  \citealt{Grankin07};  \citealt{Grankin08}; \citealt{Herbst00}).
The photometric variability  manifests   itself \rm over different timescales, from
minutes  up  to several  years  and  they  are linked  with  different
mechanisms  of  variability  (see,  e.g.,  \citealt{Messina04}).   The
photometric  variability  manifests   itself \rm in  all  photometric  bands  and
certainly  poses difficulties  when we   attempt \rm to  use color-magnitude
diagrams  (CMD) to  accurately determine  stellar ages  and  masses by
means  of   the  isochronal  fitting.  For  a   given  young  cluster,
distribution of  age of  cluster members provide  a tool to  probe the
history  of star  formation.  The  observed magnitudes  and  colors of
young   stellar    objects   are   generally    affected   by   either
fading/reddening  or  brightening/blueing  effects,  due  to  magnetic
stellar  activity,  extinction  by  circum-stellar disc  and  variable
accretion phenomena  arising from  star-disc interaction. Due  to this
the observed color-magnitude can  be very different from the intrinsic
true  values  which is  needed  for  fitting  the isochrones.   In  the
following, we show  that the longer is the  available photometric time
series the more accurate  is the determination of intrinsic magnitudes
and colors, as well as of the amplitude of variability. \\

Measurements  of I  magnitudes and  V$-$I  colors for  a fairly  large
sample of  ONC members was first carried  out by \cite{Hillenbrand97},
but from snapshot observations  and, consequently, with no information
on variability.   Time series observations  in I band  were subsequently
carried  by \cite{Herbst02}  (hereafter H02)  for ONC  members  over a
period of 45 days. Their  time series allowed them  to explore the
range  of the  I-band photometric  variability with  time  scales from
hours to weeks.  More recently, we collected five consecutive years of
I  magnitudes for a  sample of  346 ONC  members (\citealt{Parihar09};
hereafter Paper I). The improvement in measuring the average I magnitudes
and the amplitudes of variability with respect to the single season of
H02   and    with   respect   to   the    snapshot   observations   of
\cite{Hillenbrand97} is evident in  Fig.\,5 of Paper I.  The difference
between the I  band magnitude from Hillenbrand single  snapshot observations and
the mean  I band magnitude determined  by us can be  up to 2 magnitudes,
whereas difference with H02 can be up to 1 magnitude.\\

It is true  that Herbst and collaborators have  monitored ONC at
Van Vleck  Observatory (\citealt{Herbst00}) in the I  band during 1990-1999,
therefore for a time interval  much longer than ours, deriving average
magnitudes and variability of amplitudes  more accurately than us and for
a larger sample  of bright stars.  However, the major improvement  in our observations is that we could collect  simultaneously data  in three different bands
(V, R, and  I) and for 5  consecutive years. This puts us  in a better
position to derive average values to position ONC stars in CMDs and to
investigate what    impact the  photometric variability may  have on the
age and mass determination.\\ 

We  present the data in Sect.\,2 and our
analysis in Sect.\,3.  The effect of rotation and  long-term cycles on
observed magnitudes and colors are presented in Sect.\,4. The modeling
of color  variations is presented in Sect.\,5,  whereas discussion and
conclusions are presented in Sect.\,6 and 7.

\section{Data}

As  described  in  Paper  I,  we have  collected  five  seasons  (from
2003-2004  to  2007-2008)  of  I-  and  V-band  data  and  one  season
(2007-2008) of additional R-band data  for a total of 346 ONC members.
To date, this is the most  complete (in terms of number of photometric
bands) multi-year database of VRI observations for ONC members.

Our members  belong to a 10$^{\prime}\times10^{\prime}$  field of view
(FoV)   located  South-West   of   the  Trapezium   stars  and   whose
extinction-corrected magnitudes and colours  are in the range 12 $\la$
V$_0$   $\la$  20\,mag   and   0  $\la$   (V$-$I)$_0$  $\la$   5\,mag,
respectively.   Starting  from  the  2008-2009 season,  we  have  been
monitoring a larger  20$^{\prime}\times20^{\prime}$ region to make our
database more complete also in terms of members.

Our stellar  sample has been  observed with the 2-m  Himalayan Chandra
Telescope  (HCT) and  the 2.3-m  Vainu  Bappu Telescope  (VBT) of  the
Indian  Astronomical  Observatory  (IAO,  India).  On  each  telescope
pointing,  we  collected  3-4  consecutive frames  within  very  short
intervals  of time  ($<$  1\,hr) that  were  subsequently combined  to
compute  an average  magnitude  and standard  deviation. The  standard
deviation  may  be  considered   a  robust  estimate  of  the  achieved
photometric  accuracy.  More  of  such telescope  pointings were  made
during each observation  night.  We refer the reader to  Paper I for a
detailed  description  of   the  observation  strategy,  pre-reduction
process, magnitudes extraction, and data preparation.

Briefly, in order  to remove  possible outliers from  our time series  data, we
applied a  3$\sigma$ clipping  filter.  Owing to  the season-to-season
variation of  the star's average magnitude,  it turned out  to be more
accurate to  apply such a filter  to each seasonal  time series rather
than to the whole time series. The observation season generally lasted
from two to six consecutive  months and about 0.15\% observations were
discarded for the subsequent analysis. A fraction of such outliers may
arise  from intrinsic  variability,  likely related  to flare  events.
However, owing  to their very small percentage,  their exclusion does
not    affect   the    results   of    our   analysis.     In   Tables
\ref{V-photom}--\ref{V-I-photom}   we    summarize   the   photometric
properties of our targets in the V and I bands and in the V$-$I color,
respectively.  In  the columns we  list, the [PMD2009]  serial number,
number of observations, number of outliers, number of seasons, average
magnitude V$_{\rm mean}$ and its standard deviation $\sigma_{\rm tot}$
computed  for the  complete  5-yr time  series,  the average  seasonal
standard deviation $< \sigma_{\rm  seas} >$ and the standard deviation
of the seasonal mean  magnitudes $\sigma_{\rm <Vseas>}$, the brightest
and  faintest   magnitudes  in  the  complete  time   series  and  the
photometric  precision.  We  refer the  reader to  the Appendix  for a
description  of  the method  used  to  measure the  brightest/faintest
magnitudes and the amplitude of  variability.  To make a more accurate
estimate of these quantities we  considered only seasons having ten or
more observations for each star.

\begin{table*}
\caption{\label{V-photom}Summary of V-band photometry of ONC members.  In order of column, we list the [PMD2009] serial number, number of observations, of outliers, of seasons, the average magnitude V$_{\rm mean}$ and its standard deviation $\sigma_{\rm tot}$ computed for the complete 5-yr time series, the average seasonal standard deviation $< \sigma_{\rm seas} >$ and the standard deviation of the seasonal mean magnitudes $\sigma_{\rm <Vseas>}$, the brightest and faintest magnitude in the complete time series (see Appendix) and the photometric precision. The full table is available online.}
\begin{tabular}{|r|r|c|c|r|r|r|r|r|r|r|}
\hline
  \multicolumn{1}{|c|}{[PMD2009]} &
  \multicolumn{1}{c|}{\# obs.} &
  \multicolumn{1}{c|}{\# out.} &
  \multicolumn{1}{c|}{\# } &
	  \multicolumn{1}{c|}{V$_{\rm  mean}$} &
  \multicolumn{1}{c|}{$\sigma_{\rm tot}$} &
  \multicolumn{1}{c|}{$<\sigma_{\rm seas>}$} &
  \multicolumn{1}{c|}{$\sigma_{\rm <Vseas>}$} &
  \multicolumn{1}{c|}{V$_{\rm min}$} &
  \multicolumn{1}{c|}{V$_{\rm max}$} &
  \multicolumn{1}{c|}{precision} \\
    \multicolumn{1}{|c|}{number} &
  \multicolumn{1}{c|}{} &
  \multicolumn{1}{c|}{} &
  \multicolumn{1}{c|}{season} &
	  \multicolumn{1}{c|}{(mag)} &
  \multicolumn{1}{c|}{(mag)} &
  \multicolumn{1}{c|}{(mag)} &
  \multicolumn{1}{c|}{(mag)} &
  \multicolumn{1}{c|}{(mag)} &
  \multicolumn{1}{c|}{(mag)} &
  \multicolumn{1}{c|}{(mag)} \\
\hline
        1        &        1        &        0        &        1        &   19.037        &      0.0        &      0.0        &      0.0        &      0.0        &      0.0        &  0.841\\
       2        &       16        &        0        &        1        &   19.205        &     0.48        &    0.464        &    0.255        &   19.455        &   18.714        &  0.091\\
       3        &       60        &        0        &        3        &   17.868        &     0.13        &    0.101        &    0.019        &   18.123        &   17.617        &  0.039\\
       4        &      112        &        0        &        4        &   16.367        &    0.113        &    0.072        &    0.027        &   16.541        &   16.166        &   0.02\\
       5        &      231        &        0        &        4        &   16.242        &    0.253        &    0.196        &    0.071        &   16.678        &   15.858        &  0.043\\
      ...        &        ...        &    ...        &    ...        &    ...        &    ...        &    ...        &    ...        &    ...        &    ...        &    ...        \\   
         \hline
\end{tabular}
\end{table*}

\begin{table*}
\caption{\label{I-photom}Summary of I-band photometry of ONC members. The full table is available online.}
\begin{tabular}{|r|r|r|r|r|r|r|r|r|r|r|}
\hline
  \multicolumn{1}{|c|}{PMD} &
  \multicolumn{1}{c|}{\# obs.} &
  \multicolumn{1}{c|}{\# out.} &
  \multicolumn{1}{c|}{\# } &
	  \multicolumn{1}{c|}{V$_{\rm  mean}$} &
  \multicolumn{1}{c|}{$\sigma_{\rm Tot}$} &
  \multicolumn{1}{c|}{$<\sigma_{\rm seas}$} &
  \multicolumn{1}{c|}{$\sigma_{\rm Vseas}$} &
  \multicolumn{1}{c|}{Vmin} &
  \multicolumn{1}{c|}{Vmax} &
  \multicolumn{1}{c|}{precision} \\
    \multicolumn{1}{|c|}{number} &
  \multicolumn{1}{c|}{} &
  \multicolumn{1}{c|}{} &
  \multicolumn{1}{c|}{season} &
	  \multicolumn{1}{c|}{(mag)} &
  \multicolumn{1}{c|}{(mag)} &
  \multicolumn{1}{c|}{(mag)} &
  \multicolumn{1}{c|}{(mag)} &
  \multicolumn{1}{c|}{(mag)} &
  \multicolumn{1}{c|}{(mag)} &
  \multicolumn{1}{c|}{(mag)} \\
\hline
      1        &      334        &        2        &        4        &   13.944        &    0.126        &    0.099        &     0.04        &   14.283        &   13.794        &  0.026\\
       2        &      238        &        0        &        5        &   17.545        &    0.383        &    0.322        &    0.067        &   18.272        &   16.822        &  0.052\\
       3        &      320        &        0        &        5        &   15.004        &    0.038        &    0.035        &     0.01        &   15.065        &   14.931        &  0.017\\
       4        &      310        &        0        &        5        &   13.421        &    0.079        &    0.044        &    0.013        &   13.586        &   13.293        &  0.011\\
       5        &      558        &        3        &        5        &   14.318        &    0.207        &     0.19        &    0.081        &   15.078        &   13.972        &  0.029\\
    ...        &        ...        &    ...        &    ...        &    ...        &    ...        &    ...        &    ...        &    ...        &    ...        &    ...        \\   
  
     \hline\end{tabular}
\end{table*}

\begin{table*}
\caption{\label{V-I-photom}Summary of V$-$I photometry of ONC members. The full table is available online.}
\begin{tabular}{|r|r|r|r|r|r|r|r|r|r|r|}
\hline
  \multicolumn{1}{|c|}{PMD} &
  \multicolumn{1}{c|}{\# obs.} &
  \multicolumn{1}{c|}{\# out.} &
  \multicolumn{1}{c|}{\# } &
	  \multicolumn{1}{c|}{V$_{\rm  mean}$} &
  \multicolumn{1}{c|}{$\sigma_{\rm Tot}$} &
  \multicolumn{1}{c|}{$<\sigma_{\rm seas}$} &
  \multicolumn{1}{c|}{$\sigma_{\rm Vseas}$} &
  \multicolumn{1}{c|}{Vmin} &
  \multicolumn{1}{c|}{Vmax} &
  \multicolumn{1}{c|}{precision} \\
    \multicolumn{1}{|c|}{number} &
  \multicolumn{1}{c|}{} &
  \multicolumn{1}{c|}{} &
  \multicolumn{1}{c|}{season} &
	  \multicolumn{1}{c|}{(mag)} &
  \multicolumn{1}{c|}{(mag)} &
  \multicolumn{1}{c|}{(mag)} &
  \multicolumn{1}{c|}{(mag)} &
  \multicolumn{1}{c|}{(mag)} &
  \multicolumn{1}{c|}{(mag)} &
  \multicolumn{1}{c|}{(mag)} \\
\hline
       1        &        2        &        0        &        0        &    5.041        &      0.0        &      0.0        &      0.0        &    5.041        &    5.041        &  0.841\\
       2        &       36        &        0        &        1        &    1.968        &    0.532        &      0.0        &      0.0        &    3.109        &    0.891        &  0.095\\
       3        &       66        &        0        &        3        &    2.875        &    0.126        &      0.0        &      0.0        &    3.136        &    2.591        &  0.041\\
       4        &       98        &        0        &        4        &    2.932        &    0.158        &      0.0        &      0.0        &    3.162        &    2.673        &  0.022\\
       5        &      213        &        0        &        4        &    1.872        &     0.29        &      0.0        &      0.0        &    2.239        &    1.125        &   0.05\\
     ...        &        ...        &    ...        &    ...        &    ...        &    ...        &    ...        &    ...        &    ...        &    ...        &    ...        \\   
  
     \hline\end{tabular}
\end{table*}

\section{Causes of variability}

\begin{figure*}
\begin{minipage}{18cm}
\begin{center}
{\psfig{file=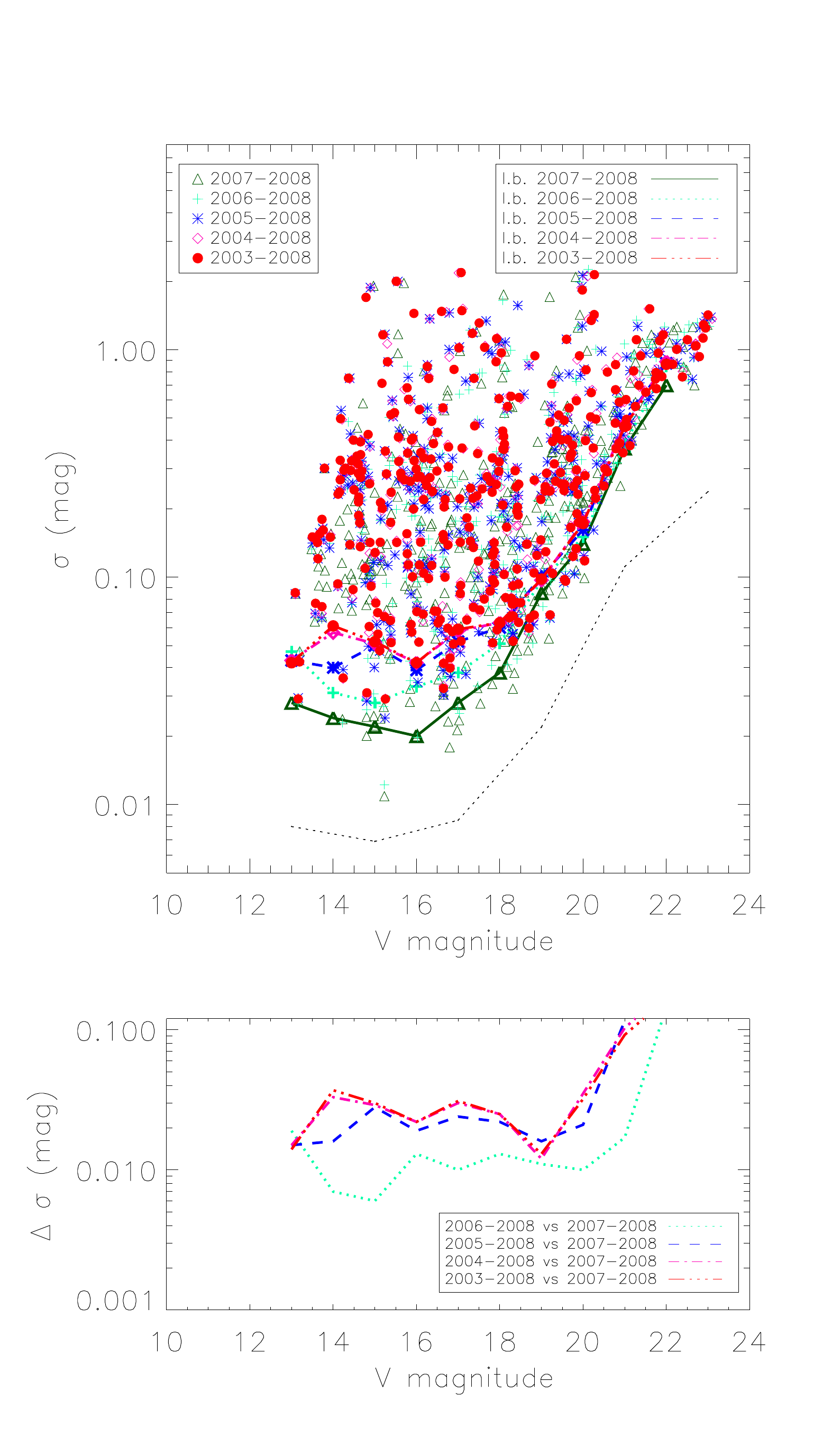,width=8cm,height=11cm,angle=0}}
{\psfig{file=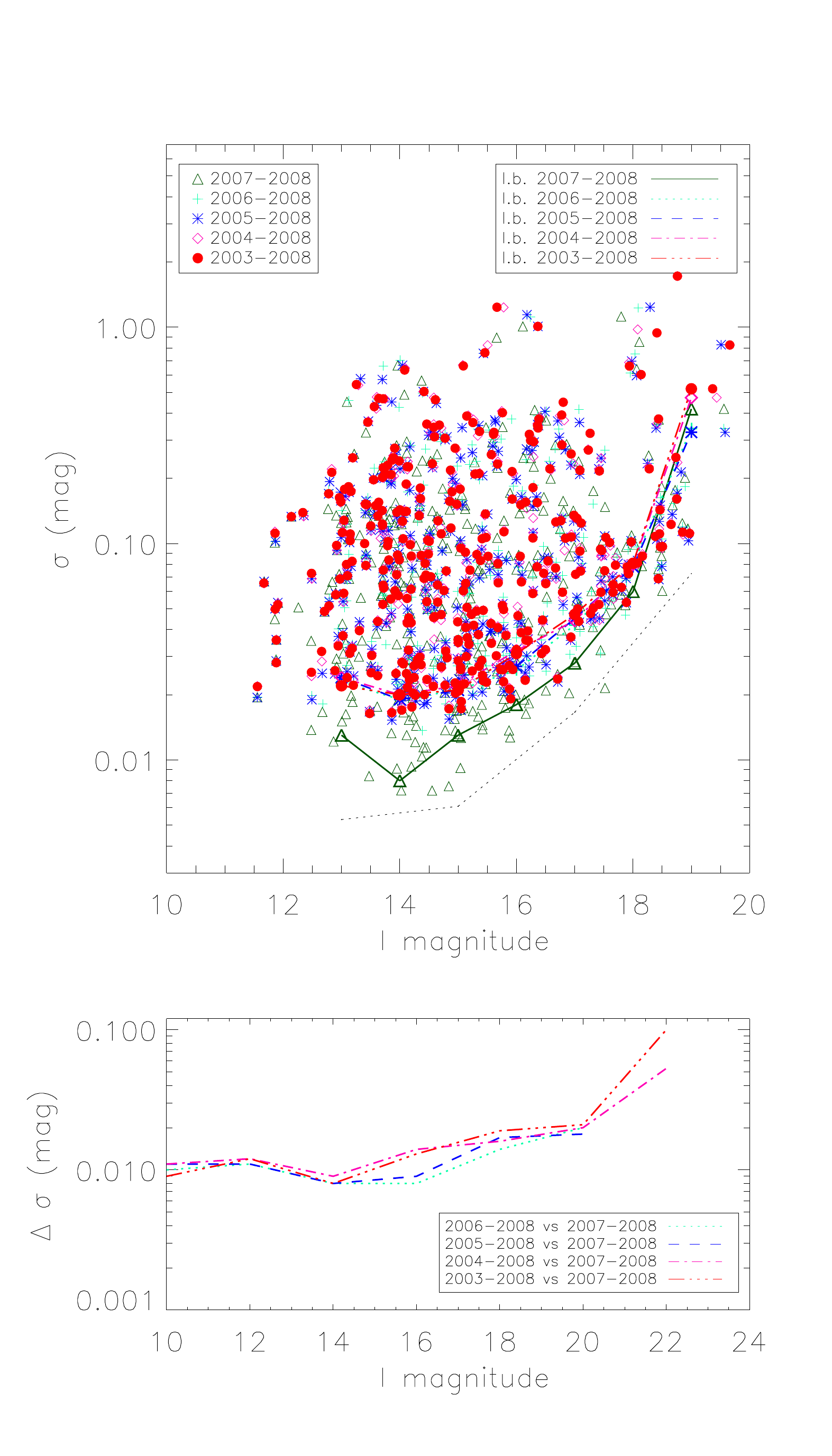,width=8cm,height=11cm,angle=0}}
\end{center}
\vspace{-0cm}
\caption{\label{rms_boundary}Top panels: standard deviation ($\sigma$) versus mean magnitude (left for V band and right for I band) for the complete sample of ONC members computed considering time intervals of increasing length. The lines represent the corresponding lower envelopes of the $\sigma$ distributions. The black dotted line represents the photometric precision. Bottom panels: differences between the lower envelopes computed considering time intervals of increasing length with respect to the last observation season (2007-2008).}
\end{minipage}
\end{figure*}

\begin{figure*}
\begin{minipage}{18cm}
\begin{center}
{\psfig{file=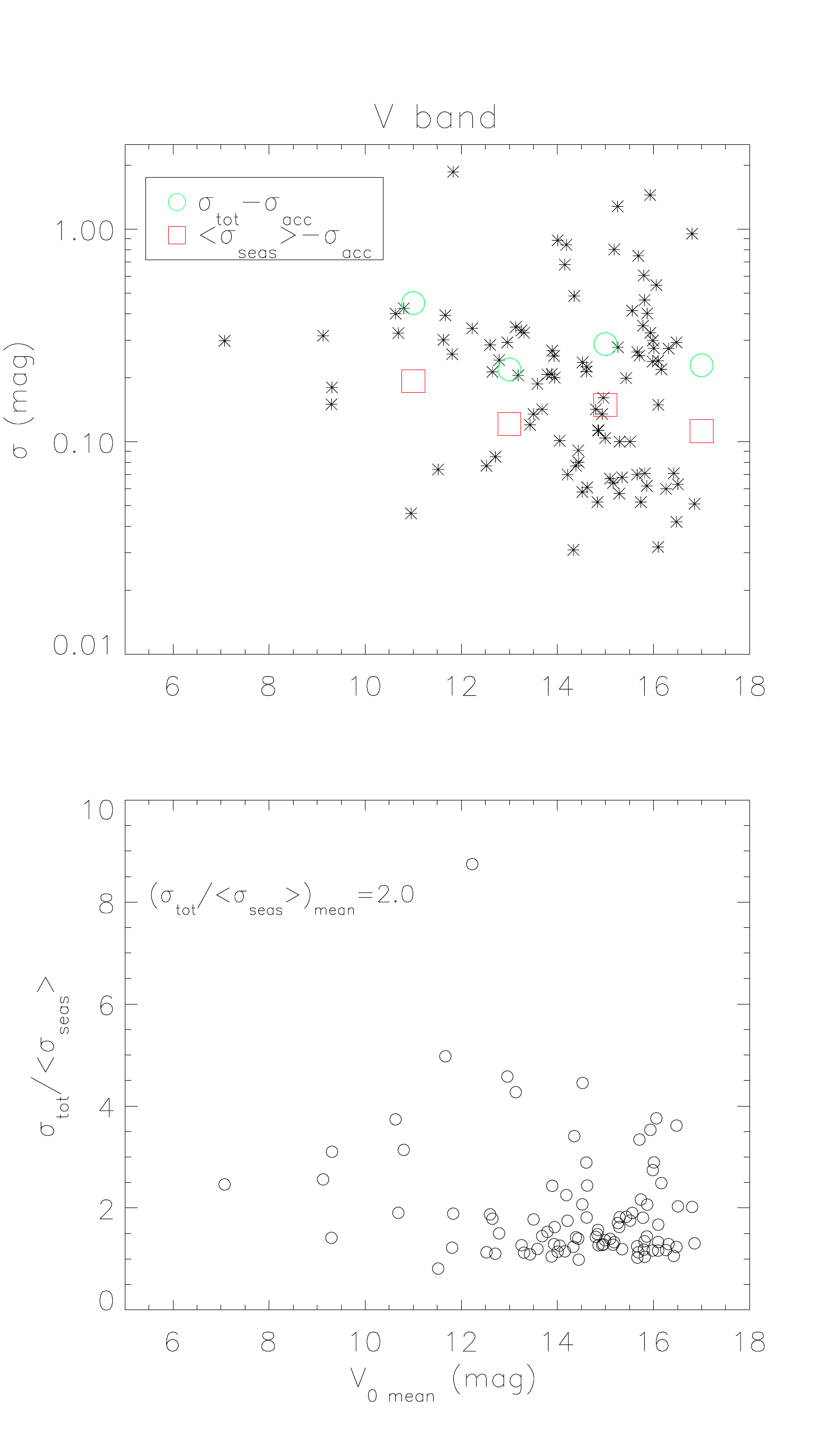,width=7cm,height=11cm,angle=0}}
{\psfig{file=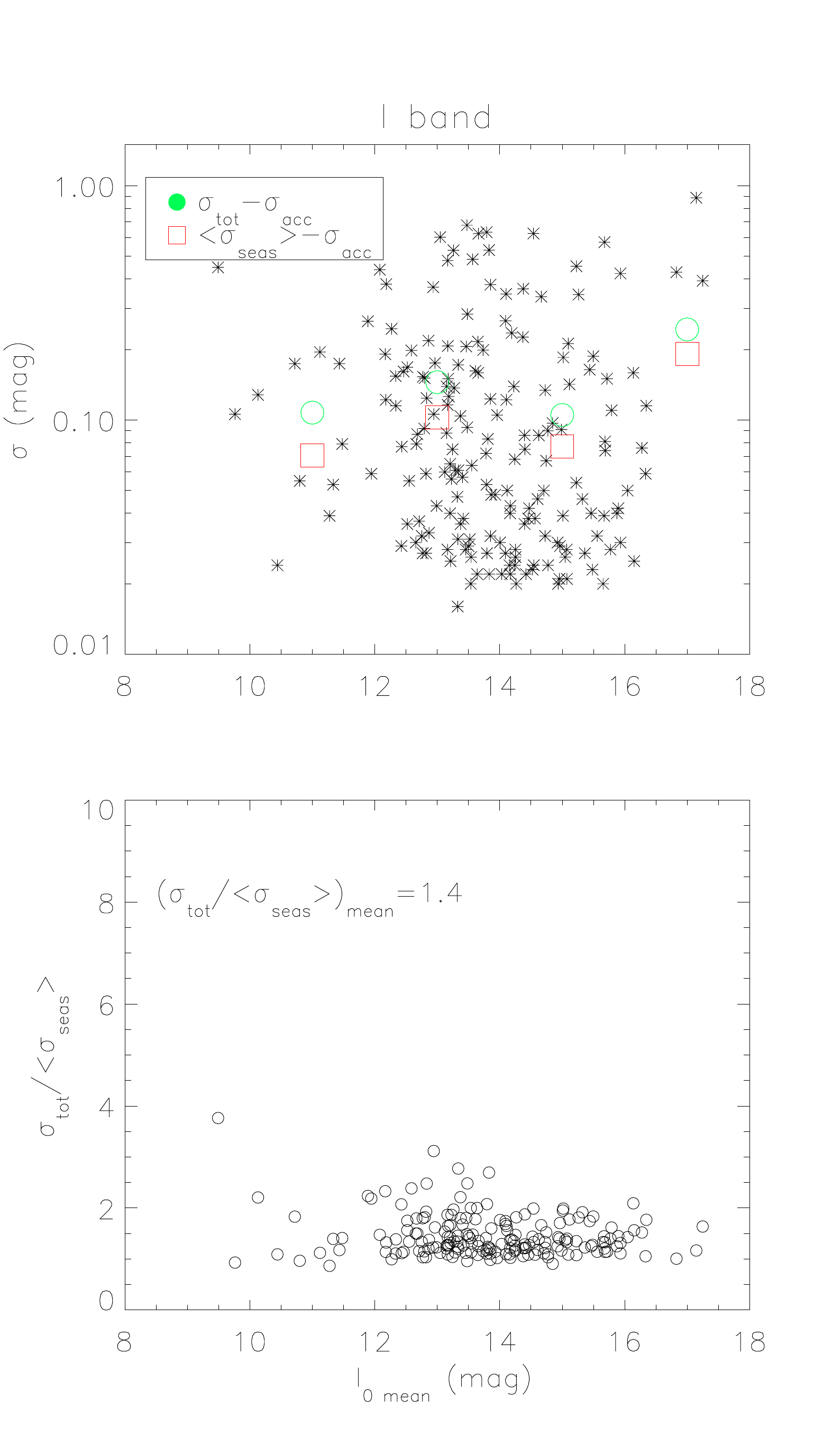,width=7cm,height=11cm,angle=0}}
\end{center}
\vspace{-0cm}
\caption{\label{rms_av}Top panels: total standard deviation  $\sigma_{\rm tot}$ (asterisks) vs. reddening-corrected mean magnitude for V (left) and I bands (right). Open circle and squares represent the average values, computed within bins of 2-mag width, of $\sigma_{\rm tot}$ and $<\sigma_{\rm seas}>$, both corrected for the photometric precision $\sigma_{\rm acc}$.  Bottom panels: ratio between  $\sigma_{\rm tot}$ and $<\sigma_{\rm seas}>$ vs. reddening-corrected mean magnitude.}
\end{minipage}
\end{figure*}

The   photometric  variability   of  low-mass   pre-main-sequence  and
main-sequence stars arises from  a number of phenomena\footnote{In our
analysis  of  variability  phenomena  we  limit to  single  stars  and
non-eclipsing binaries.} that manifest  on different time scales (see,
e.g., \citealt{Messina04}).  The phenomenon  that produces  the most
stable  and   better  characterized  variability   is  the  rotational
modulation  of  surface temperature inhomogeneities.  Its  time  scale (i.e.,  the
rotation period)  ranges from  about half a  day to several  days and,
generally, is never longer than about 20--30 days. \\

Another contribution  to the observed variability,  in which amplitude
can be larger than that arising from rotational modulation, comes from
transient  phenomena,  like  flares or  extinction-related  brightness
fadings, and manifests on shorter time scales. Owing to its stochastic
nature, it is less easily characterized.\\

Active  region  growth  and  decay  (ARGD),  variable  accretion,  and
activity cycles also contribute  to the photometric variability on the
longest time scales, from several weeks  the first to several years
the latter.  Characterization  of such variability requires multi-year
monitoring. Every  cause of variability has a spectrum of amplitudes whose range can
be accurately determined if the photometric time series is much longer
than  the characteristic   time  scales, i.e., if the   phenomenon
producing the variability is sampled over sufficiently longer periods\\

In  our study,  we can  state that  our 5-yr  long monitoring  is best
suited to characterize the  amplitude of variability arising from both
rotational modulation and  ARGD.  Whereas, observational sampling more
frequent  and uninterrupted  than ours  is required  to make  a robust
statistics of  stochastic phenomena.  To some extent  our observations
allow  us to  detect  these  phenomena, but  certainly  not  their
complete  amplitude  spectrum.   Finally,  activity cycles  and  other
long-term activities  manifest them self  over the periods from few  years to
decades.  Therefore, we can derive  only lower limits to amplitudes of
long term variability.\\

As  a matter  of  fact, the  longer  the time  series  the better  the
characterization   of   the  whole   spectrum   of  variability.    In
Fig.\,\ref{rms_boundary},  we show,  as an  example  for the  V and  I
bands,  how  the level  of  observed  variability,  and therefore  the
capability of detecting the  complete amplitude spectrum, increases as
 time series lengthens. In the same figure we  show  for  each target  the  V band  (left) and  I band  (right)
standard  deviation   versus  mean  magnitude   considering  the  last
observation  season  to  begin,  that  is  observations  collected  in
2007-2008 (green triangles), which  is our longest observation season,
and overplot its lower envelope  (green solid line).  This envelope is
computed  averaging  the  10\%  smallest  values within  each  bin  of
one-magnitude width. Then, we consider the standard deviation from the
last two years of observations (sky-blue crosses), and so on. Finally,
we consider the $\sigma_{\rm tot}$  (red bullets) of the complete time
series. \\

It  is  evident  that  as  far  as  the  series  becomes  longer,  the
variability amplitude, as measured by the standard deviation, tends to
increase.  This  happens because  we  are  obtaining  a more  accurate
estimation  of all  variability  components.  As  shown  in the  lower
panels of  Fig.\,\ref{rms_boundary}, the largest  increase is obtained
when we add  a second consecutive season of  observations, and again a
third. Starting from the fourth consecutive season, the lower envelope
continues to increase but at much  smaller rate.  In the case of the I
band observations, we see a similar behavior, although the variability
in the I band is smaller than in the V band and the standard deviation
lower envelope increases less rapidly than in the case of V band.\\

We find that  the ONC members exhibit a  photometric variability whose
average  seasonal  standard deviation  corrected  for the  photometric
accuracy    (dotted   black    lines    in   the    top   panels    of
Fig.\,\ref{rms_boundary})  is $<\sigma_{seas\_V}>  $  = 0.24\,mag  and
$<\sigma_{seas\_I}>$ = 0.12\,mag, whereas the total standard deviation
(on  a  5-yr  base  line)  is $<\sigma_{tot\_V}>  $  =  0.39\,mag  and
$<\sigma_{tot\_I}>$ = 0.17\,mag.  The average amplitude of variability
is  smaller in I  band than  in V, which  is expected  for PMS
objects.  The  use  of I  band  allows  us  to deal  with  photometric
variability a factor of 2 smaller  than V band and hence preferable in
positioning the ONC members in CMDs.\\

To investigate any dependence of the variability level on stellar mass,
we have  to correct our  mean magnitudes for  interstellar extinction.
Unfortunately,  we have values  of extinction  (A$_{\rm v}$)  for only
about  25\%  of  our   sample,  specifically  only  for  the  brighter
members.  Using  the  \citet{Hillenbrand97}  A$_{\rm  v}$  values  (see
Sect.\,3.3) to correct the magnitudes  for reddening, we find that the
average variability  levels are  $<\sigma_{tot\_V}> $ =  0.24\,mag and
$<\sigma_{seas\_V}> $ = 0.12\,mag ($<\sigma_{tot\_I}>$ = 0.12\,mag and
$<\sigma_{seas\_I}>$  =  0.09\,mag) in  the  10  $<$  V$_{\rm 0}$  $<$
15\,mag range (i.e. stars with M $>$ 0.3-0.4 M$_{\odot}$) .\\

We plot these  results in the top panels  of Fig.\,\ref{rms_av}, using
asterisks to represent $\sigma_{\rm tot}$, open bullets and squares to
represent   the  average   values   of  $\sigma_{\rm   tot}$  and   of
$<\sigma_{\rm seas}>$  (obtained using a  binning width of  2-mag). In
the bottom  panels of Fig.\,\ref{rms_av},  we plot the ratio  of these
quantities versus magnitude.  We find that the average  ratios are 2.0
and  1.4 for  the  V and  I  bands, respectively,  and no  significant
dependence with mass is found. \\

In  the following  sub-sections  we  show that  when  we consider  the
contribution to variability arising only from periodic phenomena, that
is the  light rotational  modulation due to  surface temperature inhomogeneities,  then its
impact on  mass and age estimation  is negligible. On  the other hand,
when  we consider  the additional  contribution of  other non-periodic
phenomena,  e.g., related  to  variable accretion  or  ARGD, then  the
impact  of  variability becomes  increasingly  significant.  The  best
approach  to  disentangle  the  contribution on  variability  by  
rotational  modulation  of temperature inhomogeneities to  others  is  to  analyze  separately
periodic  and non  periodic variables.

\subsection{The  samples of 'clean' and 'dirty' light curves}

One   major   difficulty   in   estimating  the   amplitude   of   the
rotation-induced variability from  time series data is the presence of
possible outliers, which can provide  incorrect average magnitude, but more specifically    incorrect    brightest   and   faintest    values,   thus
overestimating the amplitude of variability.  The contribution of ARGD
and  activity cycles  is minimized  by segmenting  the 5-yr  long time
series  into segments  each  corresponding to  the yearly  observation
season.

Periodic variables  are best suited  to identify and  remove outliers,
allowing  us to  measure  more confidently  the  correct amplitude  of
rotation induced variability.  In fact, once the   time series data of
the periodic  stars are  phased into folded  light curves,  it becomes
quite easy to  identify any data that significantly  deviates from the
trend caused    by the rotational modulation. These  outliers can arise
either  from bad measurements  or from  the above  mentioned transient
phenomena that are likely unrelated to the rotational modulation.

We know the rotation periods of 165 of the 346 stars considered in the
present analysis.  The rotation periods of 55 stars were discovered by
us (Paper I),   17 stars by \citet{Rodriguez-Ledesma09},
whereas   remaining 93  stars were reported as   periodic variables in the literature
(e.g., \citealt{Herbst00}, \citeyear{Herbst02}; \citealt{Stassun99}).

We have produced phased light  curves of all 165 periodic stars. Then,
we have automatically removed from each light curve all data deviating
more  than 3$\sigma$  from the  mean  value, and  any other  remaining
observation  that  with visual  inspection  appeared to  significantly
deviate  from   the  sinusoidal   trend  imposed  by   the  rotational
modulation.

\begin{figure}
\begin{minipage}{8cm}
\begin{center}
{\psfig{file=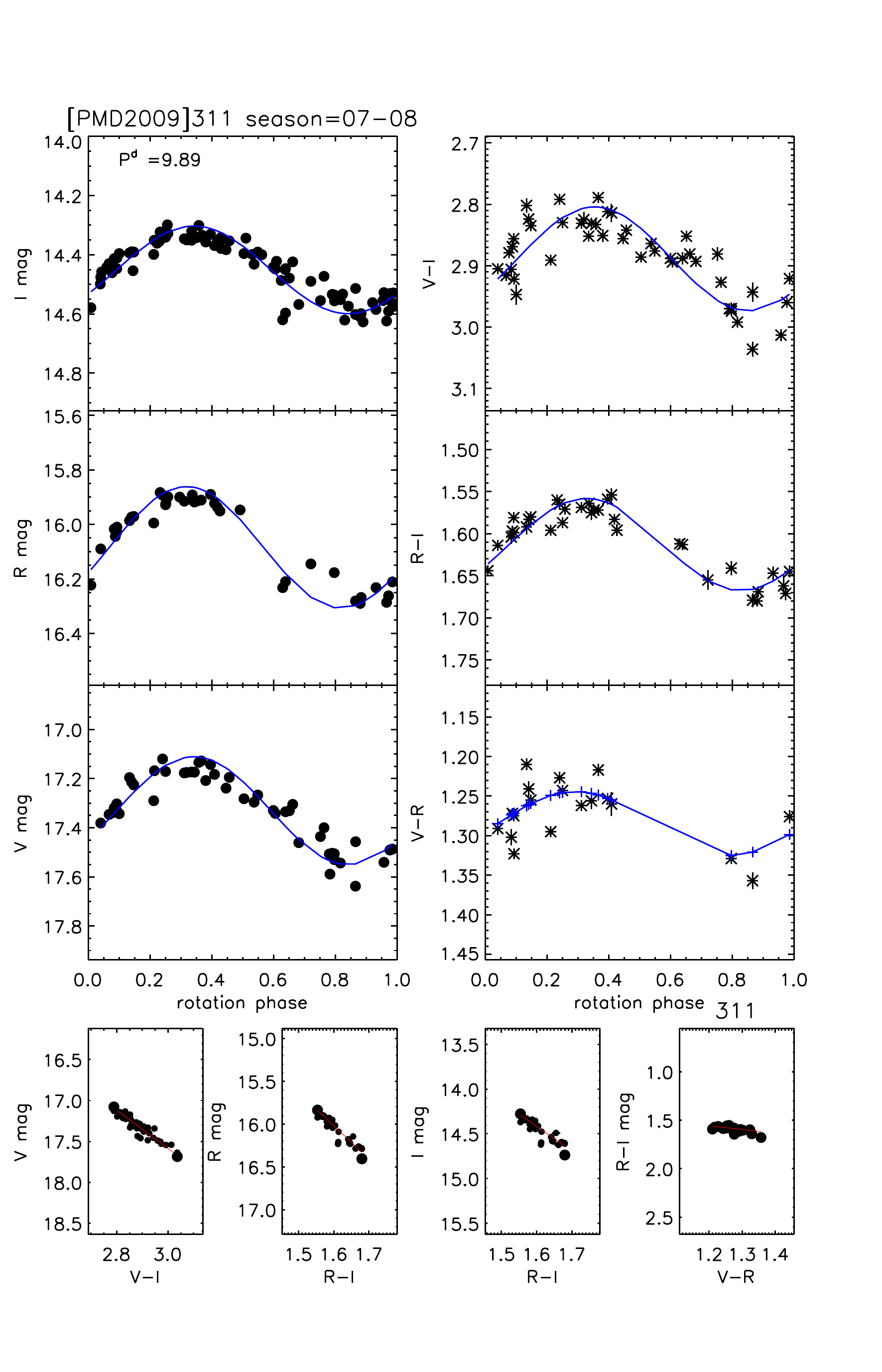,width=9cm,height=10.5cm,angle=0}}
\end{center}
\vspace{-0cm}
\caption{\label{311}Example of 'clean' magnitude and color curves (top panels) of the star [PMD2009] 311 in the season 2007-2008. 
Data are folded with the rotation period and fitted
with a single sinusoidal function (solid line). The magnitude vs. color and color vs color distributions are plotted in the botton panels
together with a linear fit (solid line).}
\end{minipage}
\end{figure}

After cleaning the phased light curves from outliers, we have selected
a subsample  consisting of very  smooth light curves. These  are light
curves that were best fitted by a single sinusoidal function and whose
ratio ({\it R})  between the amplitude  and the average  residuals from
the fit was  arbitrarily set to be $R$ $\ge$  2.5.  Since these smooth
light curves  also exhibit some  level of magnitude dispersion  at any
given  rotation phase,  in the  following we  adopt the  brightest and
faintest values  of the sinusoidal  fit as the brightest  and faintest
light curve values.  For an  example in Fig.\,\ref{311} we show a very
smooth light curve of the  star [PMD2009] 311 in the 2007-2008 season.
The  top-left  panels display  the  I, R,  and  V  magnitudes and  the
top-right  panels display  the V$-$I,  R$-$I, and  V$-$R  color curves
phased with  the rotation period P  = 9.89\,d taken from  Paper I. The
solid blue lines are the sinusoidal fits to the data with the rotation
period.  In  the bottom  panels we plot  the magnitude vs.   color and
color vs.   color distributions, which are linearly  fitted (red solid
line) and  for which the  Pearson linear correlation  coefficients and
significance  levels  are  computed.   The slope  provide  a  valuable
information  on  the  average  temperature  contrast  of  the  surface
inhomogeneities  with respect to  the unperturbed  photosphere.  These
slopes can change from star to  star, and for the same star can change
from season to season.

\begin{figure}
\begin{minipage}{8cm}
\begin{center}
{\psfig{file=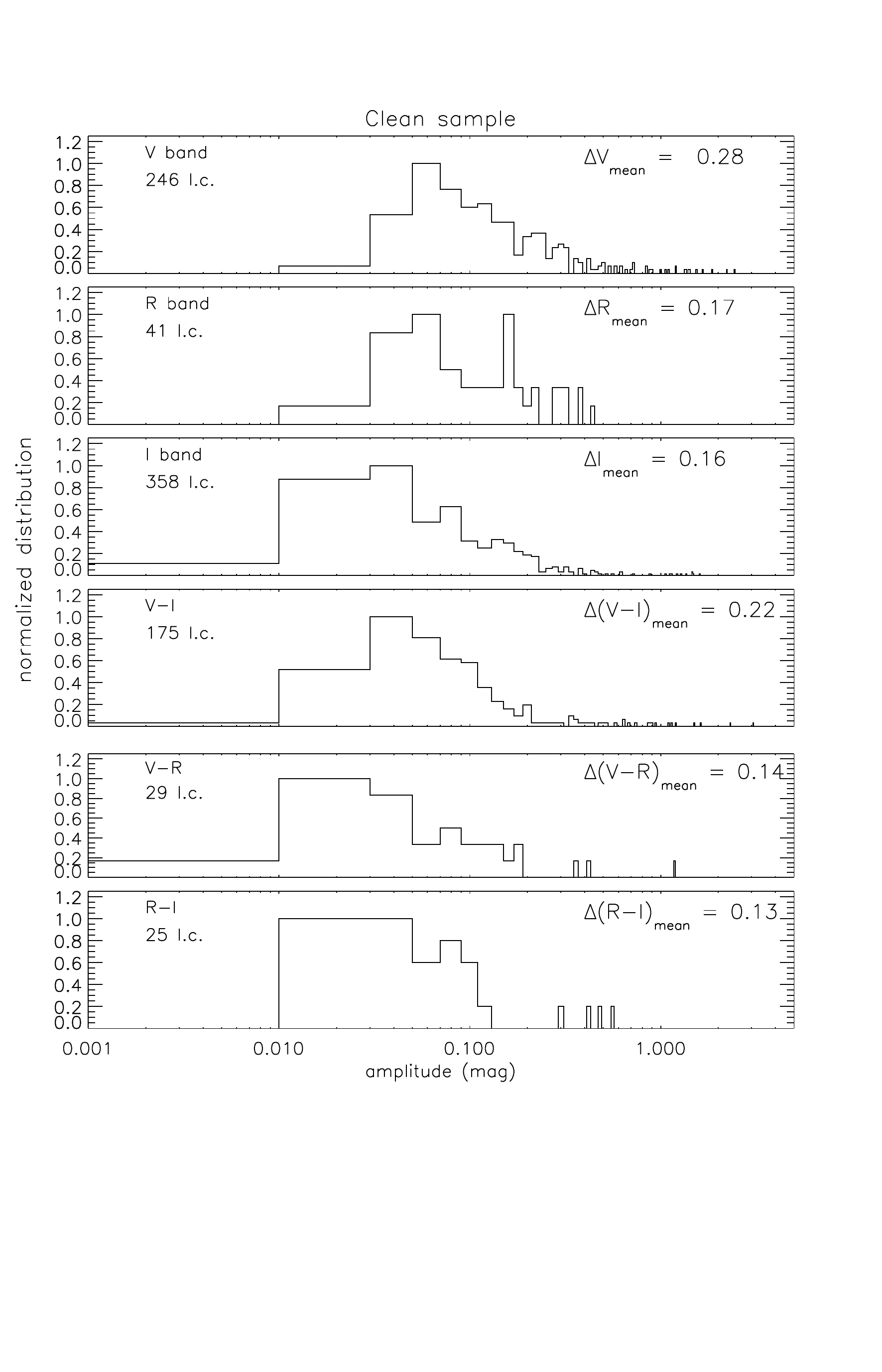,width=9cm,height=13cm,angle=0}}
\end{center}
\vspace{-3cm}
\caption{\label{histo_ampl}Normalized distributions of light and color curve  amplitudes from the clean sample. Labels indicate the number of curves (l.c.) used to build the distributions.}
\end{minipage}
\end{figure}

Using the $R$ value as  criterion to distinguish light curves, we were
left with one sub-sample of  very smooth light curves (hereafter named
'clean'  periodic) with  $R$ $\ge$  2.5  and one  sub-sample of  quite
scattered  light curves  (hereafter named  'dirty' periodic)  with 1.0
$\le R <$ 2.5. Stars with R $<$ 1, although have their rotation period
known from the literature, show  no light rotational modulation in our
time series.

We remind that  each periodic star has  up to 5 light curves  in the I
and up to 5 in the V bands (since the complete time series is analyzed
after splitting into  5 consecutive seasons), whereas it  has only one
light curve  in R band, which  is available for  the 2007-2008 season.
We find  that each  star during  the same season  can have  either all
light curves  (V, I, V$-$I)  clean or all  dirty, or a  combination in
which light  curve changed from one  season to other  season and dirty
became clean or visa versa.

In general, dirty light curves  are more numerous than the clean ones.
On a total of  797 I-band light curves, 45\% are clean;  on a total of
156 R-band  light curves,  26\% are  clean; on a  total of  800 V-band
light curves,  31\% are clean. If  we consider that about  50\% of the
complete  stellar sample  (346 stars)  is made  by non  periodic stars
(whose time  series can be considered  as dirty), then  we derive that
among ONC members clean light curves are found with a percentage never
larger than 25\%  in the I band,  and never larger than 15\%  in the V
band.  As extreme cases we  mention the following stars: [PMD2009] 57,
147,  182, 191,  256, 274,  311, 333  whose light  curves are  all and
always clean, whereas the following stars: [PMD2009] 23, 98, 100, 132,
185, 192, 224, 227, 334, 337 have all and always dirty light curves.

In Fig.\ref{histo_ampl},  we plot the normalized  distributions of the
magnitude  and color  curves amplitudes  for the  'clean' sample  on a
logarithmic scale.   Labels report the  number of light  curves (l.c.)
used to  build the distributions,  and the average  peak-to-peak light
curve  amplitude.  On  average,  the V  light  curves have  amplitudes
larger than  R and  I light  curves, and the  V$-$I color  curves have
amplitudes larger  than V$-$R  and R$-$I. Such  a trend  is consistent
with  variability  arising  from temperature  inhomogeneities  (either
hotter or cooler than the  unperturbed photosphere) carried in and out
of view  by the stellar rotation. We  note that there are  some V-band
light curves  having amplitudes larger than  0.5 mag. Generally, among pre-main-sequence weak-line T Tauri stars or main-sequence stars, whose photometric variability is dominated by cool spots, amplitudes are not larger than 0.5 mag  (see, e.g., \citealt{Messina01}, \citeyear{Messina10}, \citeyear{Messina11}). Nonetheless, exceptions exists, like the T Tauri star V410 Tau for which light curve amplitudes up to $\Delta$V = 0.7 mag are been measured (e.g., \citealt{Olah09}).\rm Our detection of larger amplitudes in a few stars then may suggest that also hot spots  on the stellar photosphere  created by
disk accretion may be present in our clean star sample.

\subsection{CMD for 'clean' periodic light curves}

To compare  our observations with  theoretical isochrones in  CMDs, we
have to remove the effect  of reddening from our measurements. We have
at our disposal the works by \citet{Hillenbrand97} and, more recently, by
\citet{DaRio10} in which the reddening of numerous ONC members is
computed.  In the  first work, we find measured  reddening A$_{\rm V}$
for 108 out  of 165 periodic stars,  in the second case for  97 out of
165. \citet{Hillenbrand97} and \citet{DaRio10} use different methods
to estimate A$_{\rm V}$, and for  our periodic stars in common to both
studies, their estimate differ on average by A$_{\rm V\_Hill}$-A$_{\rm
V\_DaRio}$=$-$0.93\,mag (with differences  ranging from $-$3.9\,mag to
1.3\,mag).   In the following analysis,  we use  only a  sub-sample of
'clean' light curves with known  extinction to build our CMDs, whereas
in Sect.\,5 we can use the  complete sample to investigate the slopes of the magnitude vs. color variations (which
are independent from  extinction).  In
order  to have  homogeneity  in the  data,  we have  used A$_{\rm  V}$
determined by Hillenbrand  only, which provided us a  larger sample of
extinction-corrected light curves.

\begin{figure*}
\includegraphics[width=100mm,height=170mm, angle=90]{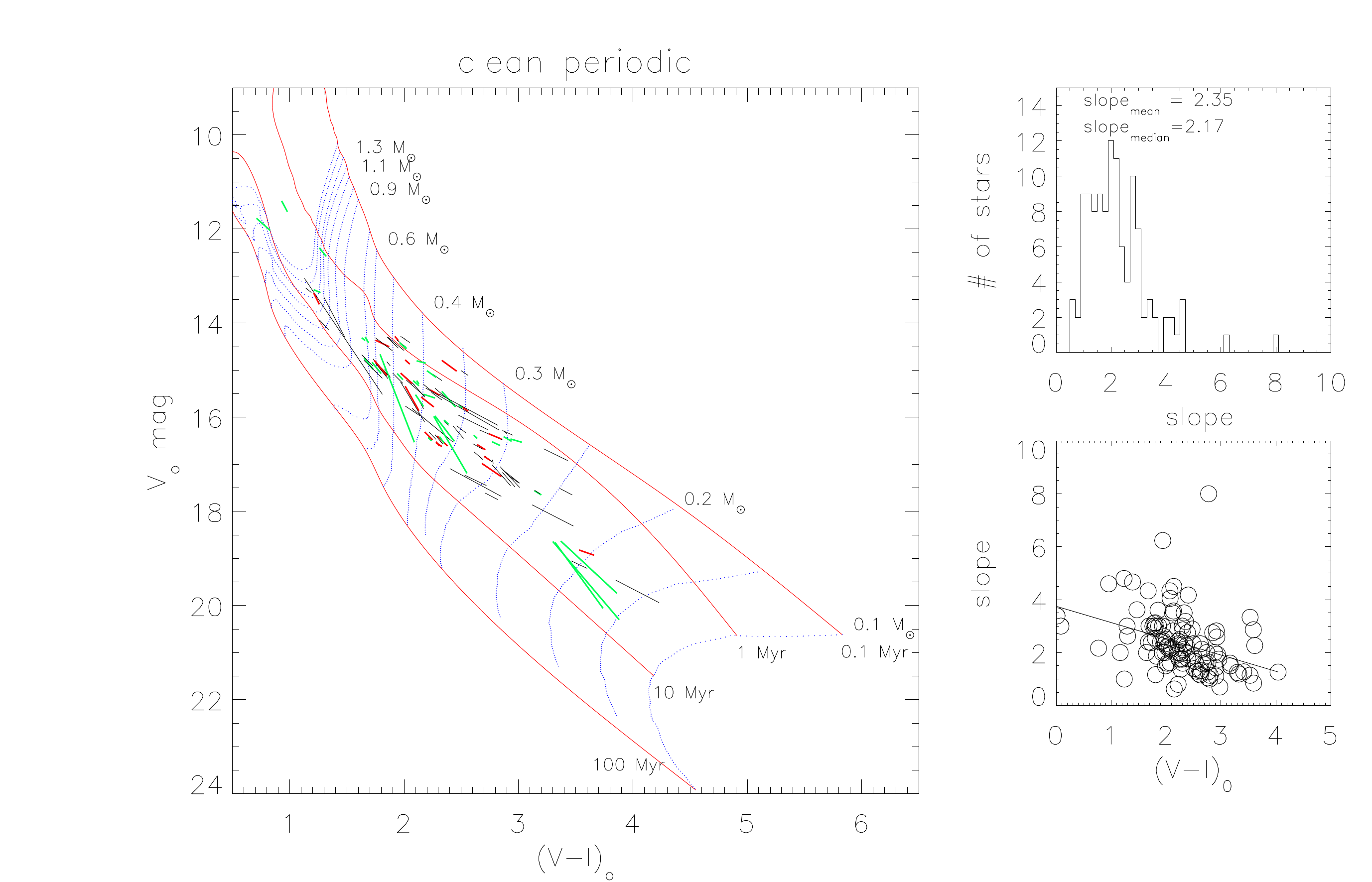}
\caption{\label{hr_vvi}Left panel: V$_0$  vs. (V$-$I)$_0$ diagram for the sample of 'clean' light curves. Red solid lines and blue dotted lines represents theoretical isochrones and mass tracks  from \citet{Siess00}. Black, green, and red segments indicate different degrees of correlation ($r > 0.95$, $0.90 < r < 0.95$, and $r < 0.90$, respectively). Top right panel: distribution of
slope of V$_0$  vs. (V$-$I)$_0$ variations, and (bottom right panel) of slope vs. (V$-$I)$_0$ color.}
\end{figure*}

In  the left  panel  of Fig.\,\ref{hr_vvi}  we  plot, in  the form  of
segments,  the   intrinsic  V$_0$  magnitude   and  (V$-$I)$_0$  color
variations derived from the clean sample of light curves. We find that
all V$_0$ vs.  (V$-$I)$_0$ variations  (line segments in
CMD)  have correlation coefficient with  significance level
larger than 95\%.  We use  black, red, and green colors to distinguish
V$_0$ vs.   (V$-$I)$_0$ variations  with correlation coefficient  $r >
0.95$, $0.90 < r < 0.95$, and $r < 0.90$, respectively.  For instance,
two or more segments may refer  to the same star, but corresponding to
different seasons.  We overplot the isochrones for 0.1, 1 and 10\,Myr,
and the zero-age-main-sequence (ZAMS; solid red lines)  computed according to the \citet{Siess00}  models for solar  metallicity. We also  plot  evolutionary
mass tracks (dotted blue lines) for different mass values in the range
of 0.1--1.5\,M$_{\odot}$.

The slopes $\Delta$V$_0$/$\Delta$(V$-$I)$_0$ exhibit a range of values
whose mean and median values  are 2.35 and 2.17, respectively (see top
right panel).   Moreover, the slopes appear to  decrease at increasing
color  (decreasing mass).   We  found similar  results  when data  are
analysed separatedly, according  to the correlation coefficient values
($r > 0.95$, $0.90 < r < 0.95$, and $r < 0.90$).

\begin{figure*}
\includegraphics[width=90mm,height=150mm, angle=90]{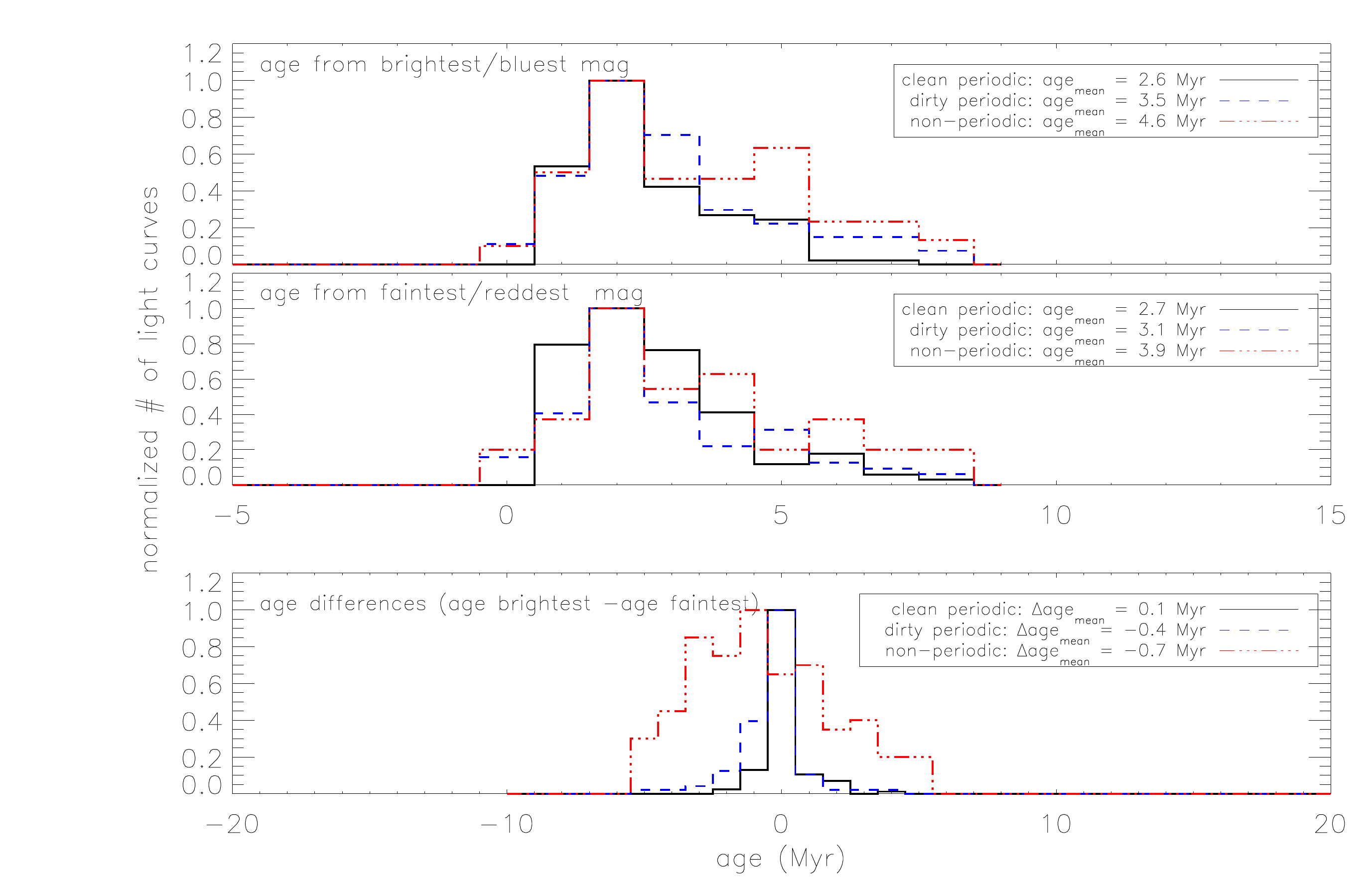}

\caption{Distribution of ages for ONC members for 'clean' period (black solid line), 'dirty' periodic (red dashed line), and non-periodic (blue dotted-dashed line)
derived from the brightest/bluest magnitude (top panel) and faintest/reddest magnitude (middle panel), and their difference (bottom panel).}
\label{age_clean}
\end{figure*}

A comparison with a family of  isochrones from \citet{Siess00} (at
s.pdf  of  1 Myr)  allows  us  to estimate  the  age  of our  targets.
Specifically, we infer  two different ages from each  light curve, one
corresponding to  the light curve brightest/bluest  value, and another
to the faintest/reddest value. The difference between the two inferred
ages  provides us  with  an  estimate of  the  impact of  variability,
arising from  rotational modulation of temperature inhomogeneities, on the age  estimate of
that star  within the same  observation season. Since, from  season to
season the  position of the targets  (represented by a  segment in the
CMD) can change owing to  AR evolution, then our results represent the
only  effects of  rotational  modulation.  From  visual inspection  of
Fig.\,\ref{hr_vvi} it  is clearly evident that on  average the presence
of surface  inhomogeneities on  the stellar surface  induces magnitude
and color variations whose slope  is similar to that of the isochrones
in  the CMD.   It means  that  variability will  not have  significant
impact on the age determination.   Nonetheless, we can compute the age
of  our  stars   and  the  results  are  plotted   in  the  panels  of
Fig.\,\ref{age_clean}.   Here,   we  plot   using   solid  lines   the
distribution of ages derived  from brightest/bluest values (top panel)
from  faintest/reddest values (middle  panel), and  their differences,
that is ages  from faintest minus ages from  brightest (bottom panel).
We find that the derived oldest and youngest mean ages are about the same (2.7
and 2.6 Myr,  respectively) and the age difference is  found to be 0.1
Myr which is not very significant. Nonetheless, we found in about 10\%
cases activity-induced  apparent age difference  is larger than  1 Myr
(up to 8 Myr).   We stress that our aim is not  to compute an absolute
age  of the  ONC members,  rather to  estimate the  age  dependency on
photometric variability. This is the  reason why our analysis does not
require different evolutionary models.

\begin{figure*}
\includegraphics[width=90mm,height=150mm, angle=90]{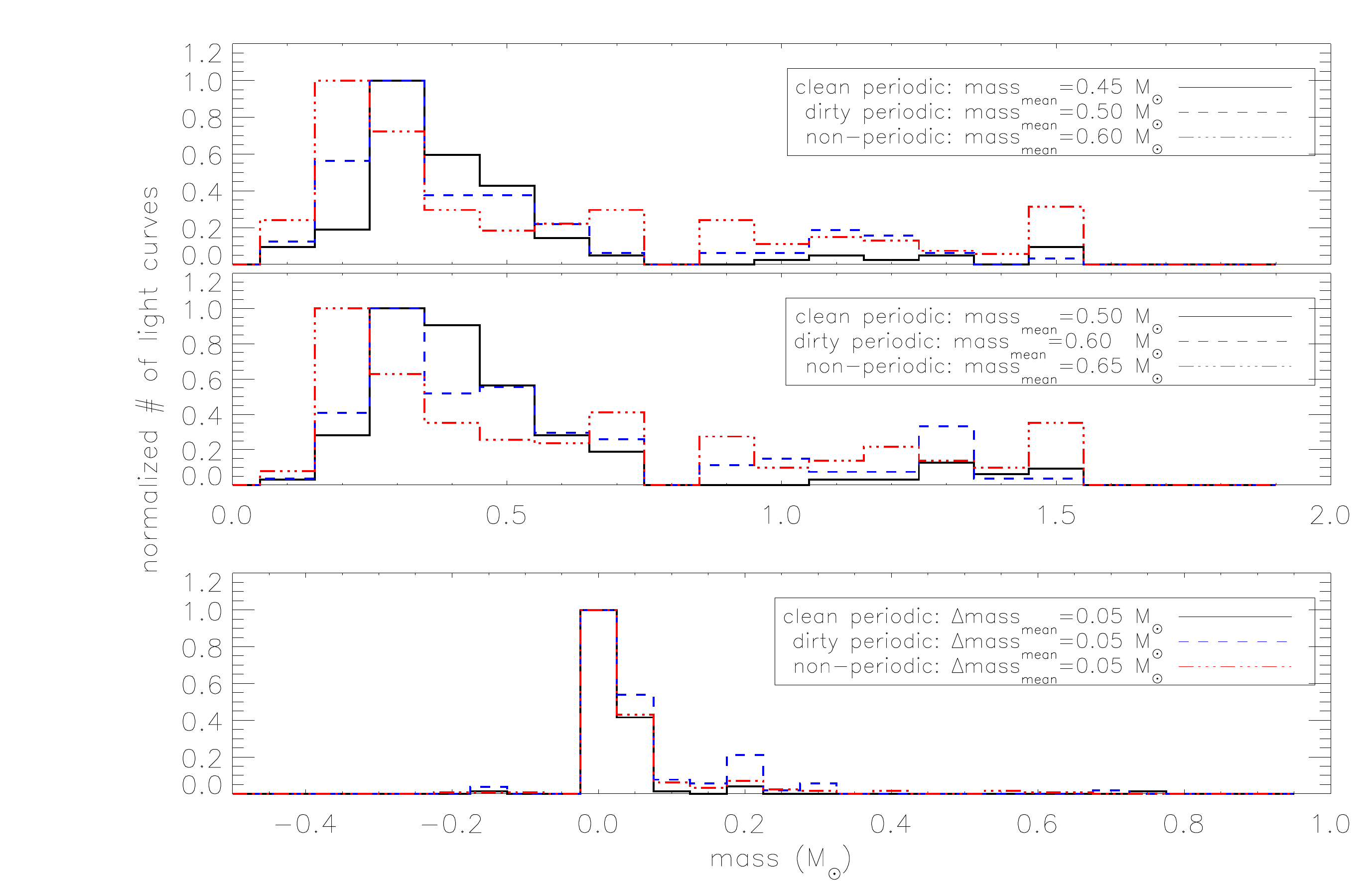}

\caption{Distribution of masses for ONC members for 'clean' period (black solid line), 'dirty' periodic (blue dashed line), and non-periodic (red dotted-dashed line)
derived from the brightest/bluest magnitude (top panel) and faintest/reddest magnitude (middle panel), and their difference (bottom panel).}
\label{mass}
\end{figure*}

\begin{figure*}
\includegraphics[width=100mm,height=170mm, angle=90]{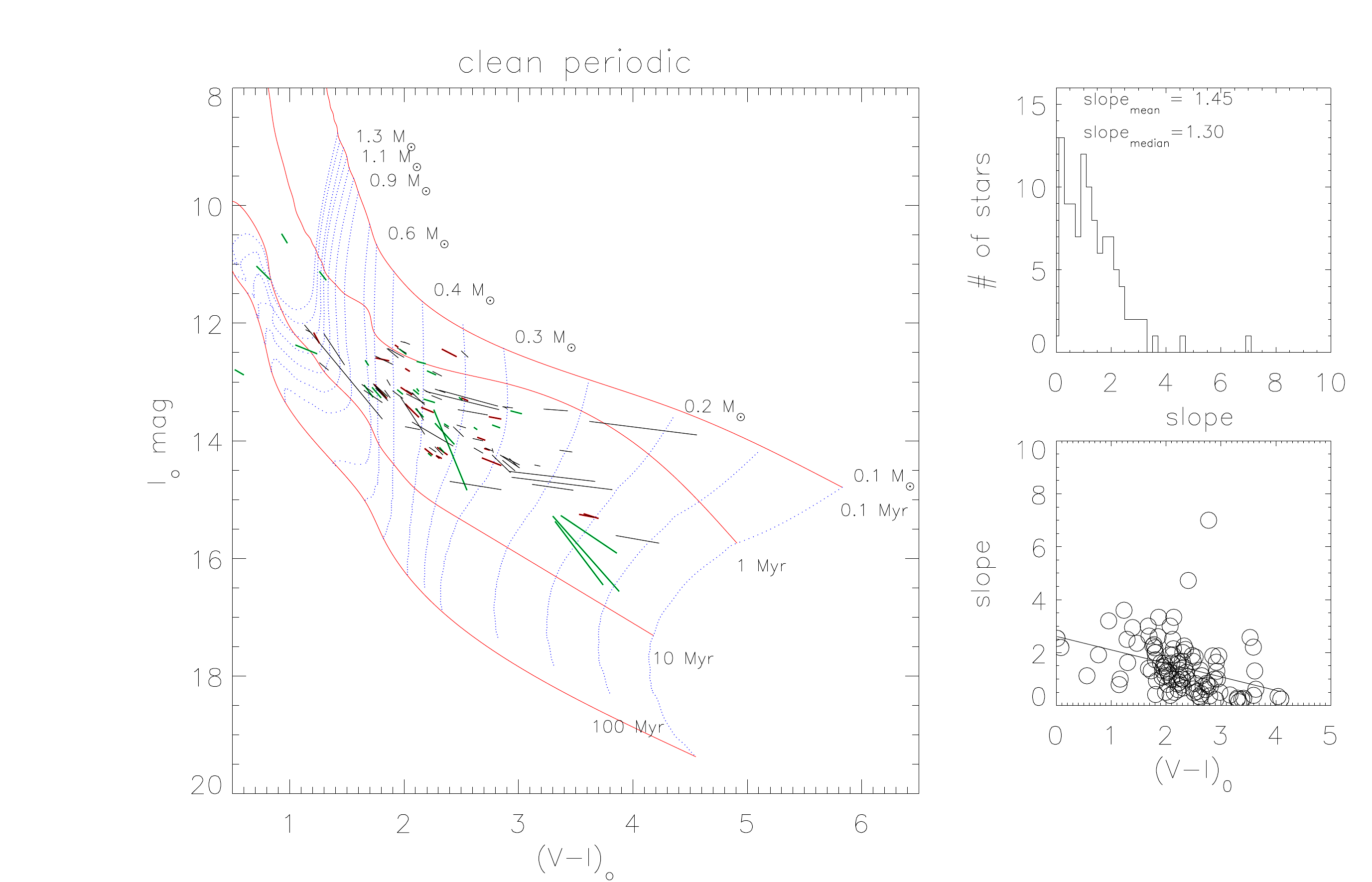}
\caption{Same as in Fig.\,\ref{hr_vvi}, but for I$_0$ vs. (V$-$I)$_0$}
\label{hr_ivi}
\end{figure*}

Similarly, we can derive the mass  of our targets by comparison with a
family  of mass  evolutionary tracks  (at s.pdf  of 0.1\,M$_{\odot}$).
Again, we derive two mass values for each light curve corresponding to
the brightest/bluest and the faintest/reddest values.  The results are
plotted in Fig.\,\ref{mass}, where  we show the distribution of masses
derived  from brightest/bluest  (top panel),  faintest/reddest (middle
panel), and their difference (bottom panel).  The derived mean values
of stellar mass are  found to be 0.46\,M$_\odot$, 0.50\,M$_\odot$, and
0.04\,M$_\odot$, respectively.  Again, the difference  in the derived
masses   are  smaller   than   the  precision   associated  with   the
determination  of mass  from the  CMD (i.e.,  the  0.1M$_\odot$ step).
Nonetheless, in about 4\% of  cases the mass difference is larger than
0.1 M$_{\odot}$.   We note in Fig.\,\ref{hr_vvi}  that the sensitivity
of mass versus V$-$I color  variations decreases as we go toward lower
masses:  a mass  difference of  0.1\,M$_{\odot}$ is  accompanied  by a
color variation  of $\Delta$(V$-$I) = 0.01\,mag at  M = 1\,M$_{\odot}$
and     $\Delta$(V$-$I)     =     0.1\,mag     at    about     M     =
0.2\,M$_{\odot}$. Therefore,  although the bottom right  region of the
CMD is populated by  longer segments, the corresponding mass variation
is not  larger at all.\\ We  have carried out a  similar analysis (see
Fig.\,\ref{hr_ivi}) for the intrinsic I$_0$ vs. (V$-$I)$_0$ variations
and as  expected obtained similar results.  Likewise previous finding,
the activity-induced apparent age differences derived in 10\% of light
curves are larger than 1\,Myr.

We conclude  that if  all ONC members  were periodic variable  and all
light curves  smooth, then, the  impact of photometric  variability on
the determination of  average age would be negligible,  and CMDs would
be well suited  to infer both mass and age. Only  a fraction (10\%) of
stars  may have  their  age  and mass  significantly  affected by  the
photometric variability.  However, as  already mentioned only  50\% of
ONC stars in our small sample are periodic and in total only 25\% have
smooth light curves.  Therefore, it is important to  see what role the
variability plays in the remaining targets.

\subsection{CMD for 'dirty' periodic  light curves}

We make an attempt to extend our analysis to the periodic  'dirty' sample. Owing to the large scatter of the phased light curves
we have taken more  care to measure the brightest/faintest values and, therefore,  the variability amplitude.
In the Appendix, we provide a detailed discussion on the approach adopted  to measure these quantities.

\begin{figure*}
\includegraphics[width=90mm,height=160mm, angle=90]{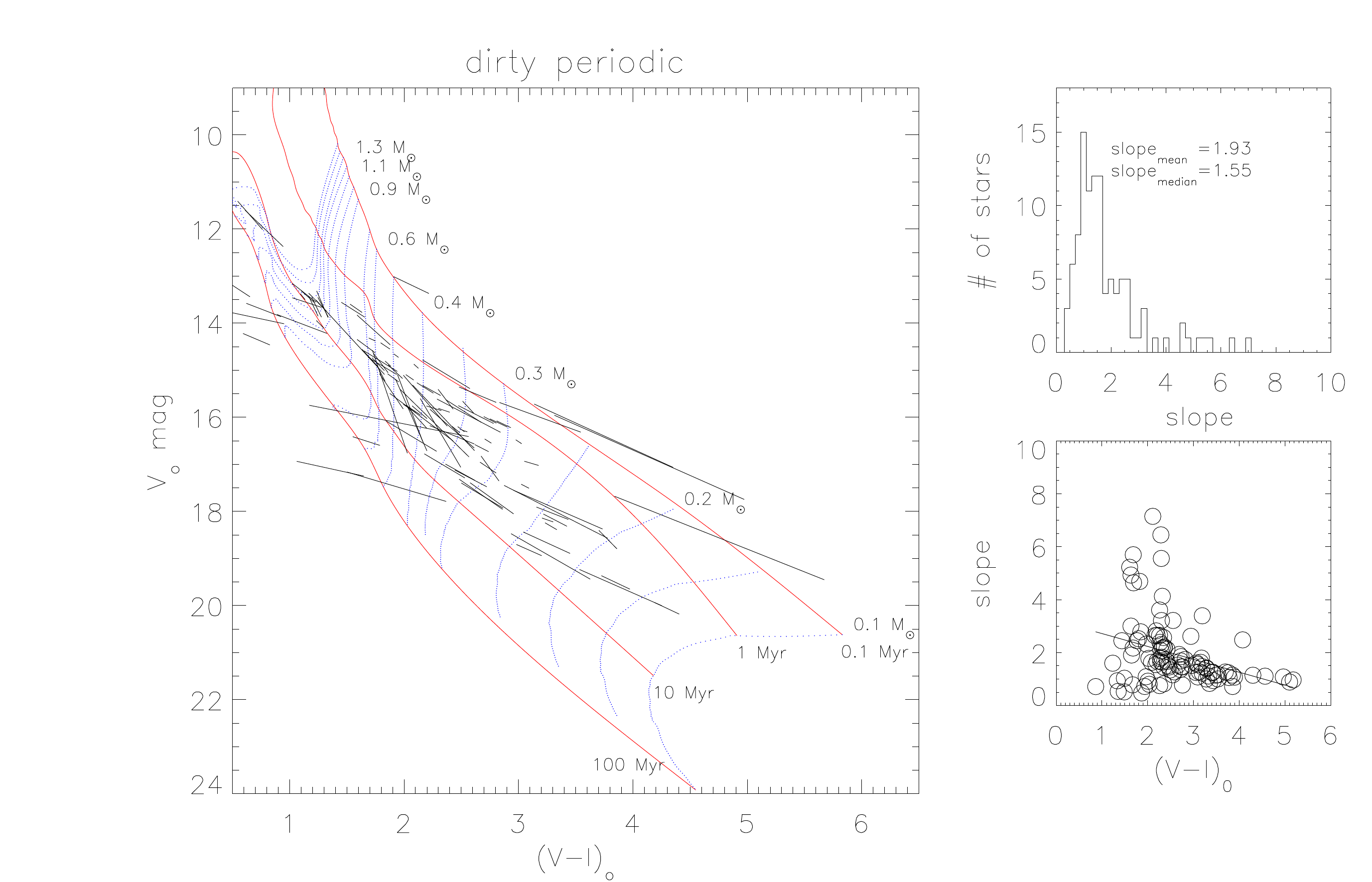}
\caption{\label{hr_vvi_sig}Left panel: V$_0$  vs. (V$-$I)$_0$ diagram for the sample of 'dirty' light curves. Top right panel: distribution of
slope of V$_0$  vs. (V$-$I)$_0$ variations, and (bottom right panel) vs. (V$-$I)$_0$ color.}
\end{figure*}

Once determined the amplitude of the periodic 'dirty' light curves,  we  can build  a CMD  and use  it to
estimate the ages and masses corresponding to the brightest/bluest and
faintest/reddest    values, as
we carried  out for 'clean'  sample.     The    results    are    plotted    in
Fig.\,\ref{hr_vvi_sig}, where we have selected only light curves where
the magnitude and color  variations are correlated with a significance
level larger than 90\%.\\

We note a few relevant differences.  First, all stars have correlation
coefficients $r < 0.90$. Second,  a number of stars appear in regions
of the CMD where no stars are expected, that is a few stars are either
too blue to be PMS stars or too red to be older than about 1\,Myr.
Finally, the  average slope of  the V$_0$ vs. (V$-$I)$_0$  relation is
smaller  than  in  the  'clean'  sample,  and  has  an  average  value
$<\Delta$V$_0$/$\Delta$(V$-$I)$_0$$>$  = 1.9  (see top-right  panel of
Fig.\,\ref{hr_vvi_sig}).   This means  that the  dirty  periodic stars
seem to  exhibit either larger  color variations or  smaller magnitude
variations  with respect to  clean periodic  stars. Since  the average
V$_0$  vs. (V$-$I)$_0$ variation  is less  steep than  the isochrones,
therefore, age derived from  the  faintest  value  will make star  to appear younger  than age derived from the brightest value.\\

 In  our  analysis, if  we also  consider   the  light  curves  whose
correlation coefficients have significance  level smaller than 90\%,  we
find     that     the     average     slope    is     even     smaller
($<\Delta$V$_0$/$\Delta$(V$-$I)$_0$$>$ = 1.35).\\

As  done  for  'clean'  light  curves,  in  Fig.\,\ref{age_clean}  and
Fig.\,\ref{mass}, we plot  the distribution of
ages and  masses inferred from  the CMD by  comparing brightest/bluest
and  faintest/reddest  values with  isochrones  and evolutionary  mass
tracks, using  dashed blue lines.  We find 10\% stars with age older than 10\,Myr, and excluding
these stars, the average age  is 3.5\,Myr and 3.1\,Myr (from brightest
and faintest magnitudes, respectively). The average difference between
the age  from brightest and faintest is  still insignificant, although
larger than in the case  of 'clean' stars, that is $-$0.4\,Myr against
0.1\,Myr.  However, the percentage of stars with age difference larger
than  1 Myr has  increased up  to 20\%  (with respect  to the  10\% of
'clean' stars).  Therefore, when  we consider the dominant variability
component  due  to  temperature inhomogeneities   together  with  the  additional  secondary
component arising from either variable accretion or ARGD or stochastic
events like  micro-flares, then  the average age  appears to  be older
(3.2\,Myr against  2.7\,Myr of  clean periodic stars)  as well  as the
percentage of  stars whose age  estimate is significantly  affected by
the variability  increases up to 20\%.  The average age  is still older
when we consider stars that exhibit periodic light curves but with not
much  tight   correlation  between  magnitude   and  color  variations
(significance level smaller than  90\%).  Concerning the mass, we find
the   average   values   0.5\,M$_\odot$   and   0.60\,M$_\odot$   from
brightest/bluest    magnitude    and   faintest/reddest    magnitudes,
respectively,  and an average  difference of  0.04\,M$_\odot$.  Again,
this differences are smaller than the precision associated to our mass
determination.

\subsection{CMD for dirty non-periodic stars}

\begin{figure*}
\includegraphics[width=90mm,height=160mm, angle=90]{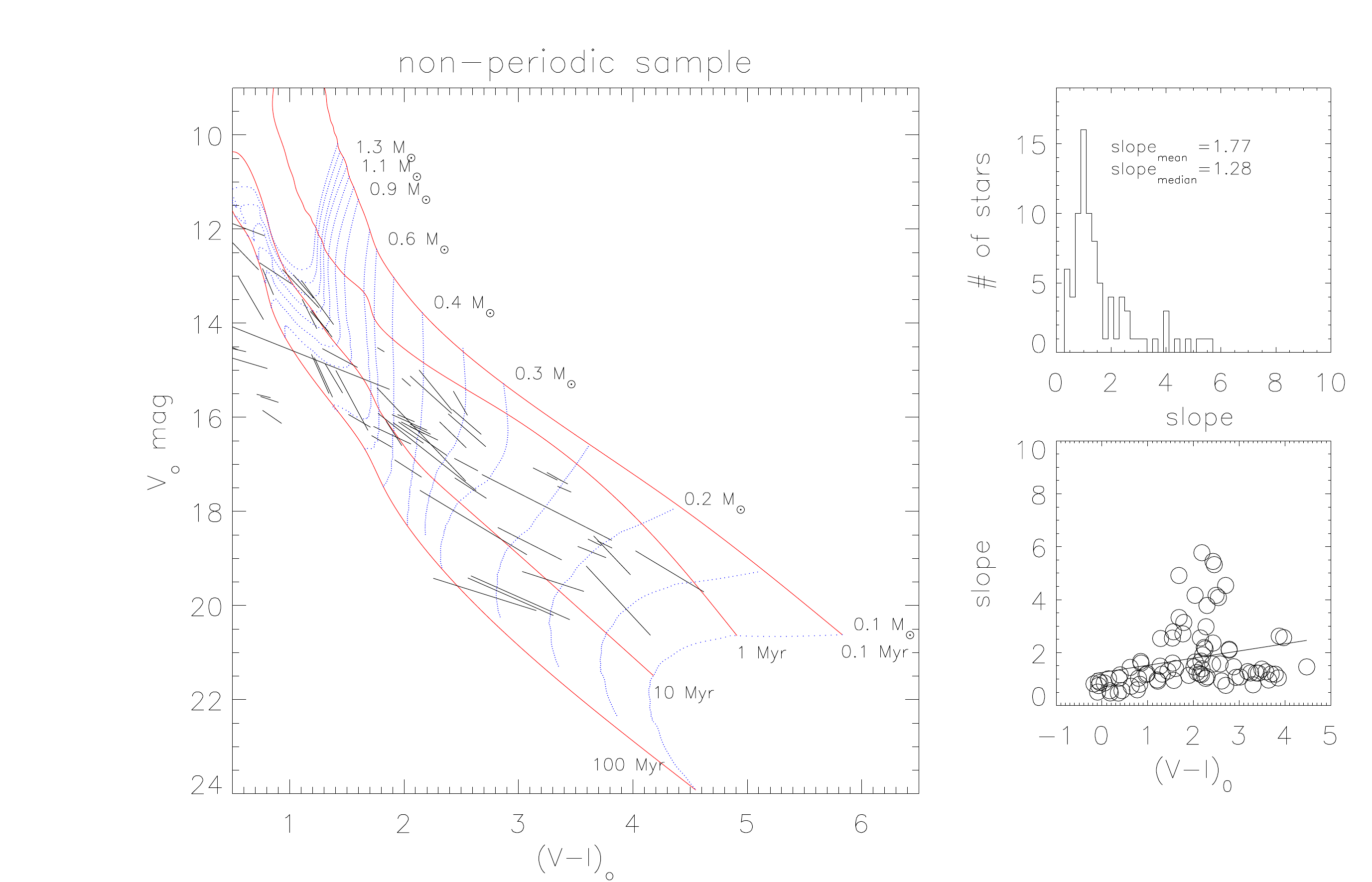}

\caption{Left panel: V$_0$  vs. (V$-$I)$_0$ diagram for the sample of non-periodic stars. Top right panel: distribution of
slope of V$_0$  vs. (V$-$I)$_0$ variations, and (bottom right panel) vs. (V$-$I)$_0$ color.}
\label{hr_vvi_sig_tot}
\end{figure*}

We  consider now  the sample  consisting of  non periodic  stars whose
light  curves   can  be  assimilated  to  'dirty'   light  curves.  In
Fig.\,\ref{hr_vvi_sig_tot} we plot the color-magnitude diagram as well
as the distribution  of slopes and their trend  with (V$-$I)$_0$ color
as done in the previous cases.  We see that the average slope of V$_0$
vs. (V$-$I)$_0$ is $<\Delta$V$_0$/$\Delta$(V$-$I)$_0>$ = 1.48, similar
to that one obtained  for  'dirty' stars with un-correlated color/magnitude
variations.   In  Fig\,\ref{age_clean}, we  plot  the distribution  of
inferred ages. Excluding  stars older than 10 Myr,  which now represent
about  50\%  of the  sample,  we find  that  the  age from  brightest
magnitude is 4.6\,Myr, from faintest magnitudes is 3.9\,Myr, and their
difference is  now $-$0.7\,mag.  \\ 

Concerning  the mass, we  find the  average values  0.6\,M$_\odot$ and
0.65\,M$_\odot$  from brightest/bluest magnitude  and faintest/reddest
magnitudes,   respectively,   and  an   average   difference  of   and
0.05\,M$_\odot$.   Again,   the  differences  are   smaller  than  the
precision associated to our  mass determination.  From our analysis we
found that the average age  of stars progressively increases from stars
showing  photometric variability,  clean periodic  (primarily  due to
cool  spots) to  dirty  periodic (due to a combination of cool  and hot  spots)  to
'irregular' photometric variability (dominated by accretion/stochastic
phenomena).   Furthermore, we  found the irregular  variables to  have different
ages derived  from brightest  and faintest magnitudes,  whereas clean
periodic star  age is  unaffected by variability.   We can  conclude by
stating that,  assuming all members are  coeval, irregular photometric
variability makes the  stars to be appeared older  and
derived ages increase with amplitude of variability.

\section{Rotation + longterm  variability}

The analysis carried out in the previous sections considers the effect
of seasonal photometric variability on
determination of stellar age and the mass arising from rotational modulation
of  light  due  to  temperature inhomogeneities  plus any  additional  random  component.  However, owing to ARGD and
activity cycles the  brightest and faintest light curve  values of the
same  star can  change  from  season to  season.   Therefore, in  this
section  we have  selected those  clean stars  which have  smooth light
curves over  at least two  observation seasons and have  plotted their
magnitudes  and  the colors  in  the  CMD  as segment  connecting  the
brightest/bluest and  faintest/reddest values ever  observed.  In this
way,  we consider  the additional  component in  amplitude  related to
long-term variability.  However, limited data set collected over only 5
years may not  be enough to capture the  complete range of variability
linked  with fairly long-term  trend.  Therefore,  our finding  on the
impact of variability  in the age and mass  determination may be taken
with some caution.

 We  find  that  the  average  difference  between  ages  derived  from
brightest  and  faintest  values  is  about zero.  However,  the  most
relevant result is that the  fraction of stars with differences larger
that  1  Myr  is  now   increased  from  10\%  (when  only  rotational
variability was considered) to about 30\%.

\section{Slope modelling}

An  important  information that  we  can  derive  from our  multi-band
monitoring  is the  dependency of  magnitude versus  color variations,
which depends  on the  contrast of temperature  between the  spotted  and
the  surrounding  unspotted  photosphere.  Here,  again  we  analyze
separately  the clean  and dirty  samples. In  the top-right  panel of
Fig.\,\ref{hr_vvi}, we  plot the distributions of  slopes derived from
the linear  fit of V magnitude  versus V$-$I color  variations for the
clean sample; whereas in top-right panel of Fig.\,\ref{hr_vvi_sig} and
of  Fig.\,\ref{hr_vvi_sig_tot} we plot  similar distributions  for the
dirty periodic  and non periodic  samples. We  note that the average derived slopes for 
the dirty periodic light curves (2.35)   and  irregular variables (1.77) are smaller than  the slopes derived for clean periodic stars. 

\begin{table*}
\caption{\label{correl}Summary of slopes and correlation coefficients of magnitude and color variation among different photometric bands. The full table is available online.}
\begin{tabular}{|rcccrcrcrcrcccrcccrcccrcrcrc}
\hline
  \multicolumn{1}{|c|}{[PMD2009]} &
  \multicolumn{1}{c|}{season} &
  \multicolumn{1}{c|}{HJD$_{\rm mean}$} &
  \multicolumn{1}{c|}{N$_{\rm I}$} &
  \multicolumn{1}{c|}{N$_{\rm R}$} &
  \multicolumn{1}{c|}{N$_{\rm V}$} &
  \multicolumn{1}{c|}{b$_{\rm V\_VI}$} &
  \multicolumn{1}{c|}{r$_{\rm V\_VI}$} &
  \multicolumn{1}{c|}{b$_{\rm R\_RI}$} &
  \multicolumn{1}{c|}{r$_{\rm R\_RI}$} &
  \multicolumn{1}{c|}{b$_{\rm V\_VR}$} &
  \multicolumn{1}{c|}{r$_{\rm V\_VR}$} &
  \multicolumn{1}{c|}{b$_{\rm RI\_VR}$} &
  \multicolumn{1}{c|}{r$_{\rm RI\_VR}$} \\
\hline
     ...        &        ...        &    ...        &    ...        &    ...        &    ...        &    ...        &    ...        &    ...        &    ...        &    ...  & ... & .... & ....      \\   
  22 & 06-07 & 2454443.8013 & 76 & 33 & 38 & 2.748 & 0.486 & 1.777 & 0.895 & 0.371 & 0.824 & -0.581 & 0.57\\
  25 & 06-07 & 2454443.8013 & 77 & 34 & 43 & 0.996 & 0.984 & 0.908 & 0.708 & 0.982 & 0.994 & -0.014 & 0.627\\
  26 & 06-07 & 2454443.8013 & 79 & 34 & 39 & 0.938 & 0.96 & -0.530 & 0.665 & 0.966 & 0.994 & 0.226 & 0.64\\
     ...        &        ...        &    ...        &    ...        &    ...        &    ...        &    ...        &    ...        &    ...        &    ...        &    ...  & ... & .... & ....      \\   
   \hline
   \end{tabular}
\end{table*}

\begin{figure}
\begin{minipage}{10cm}
\centerline{
\psfig{file=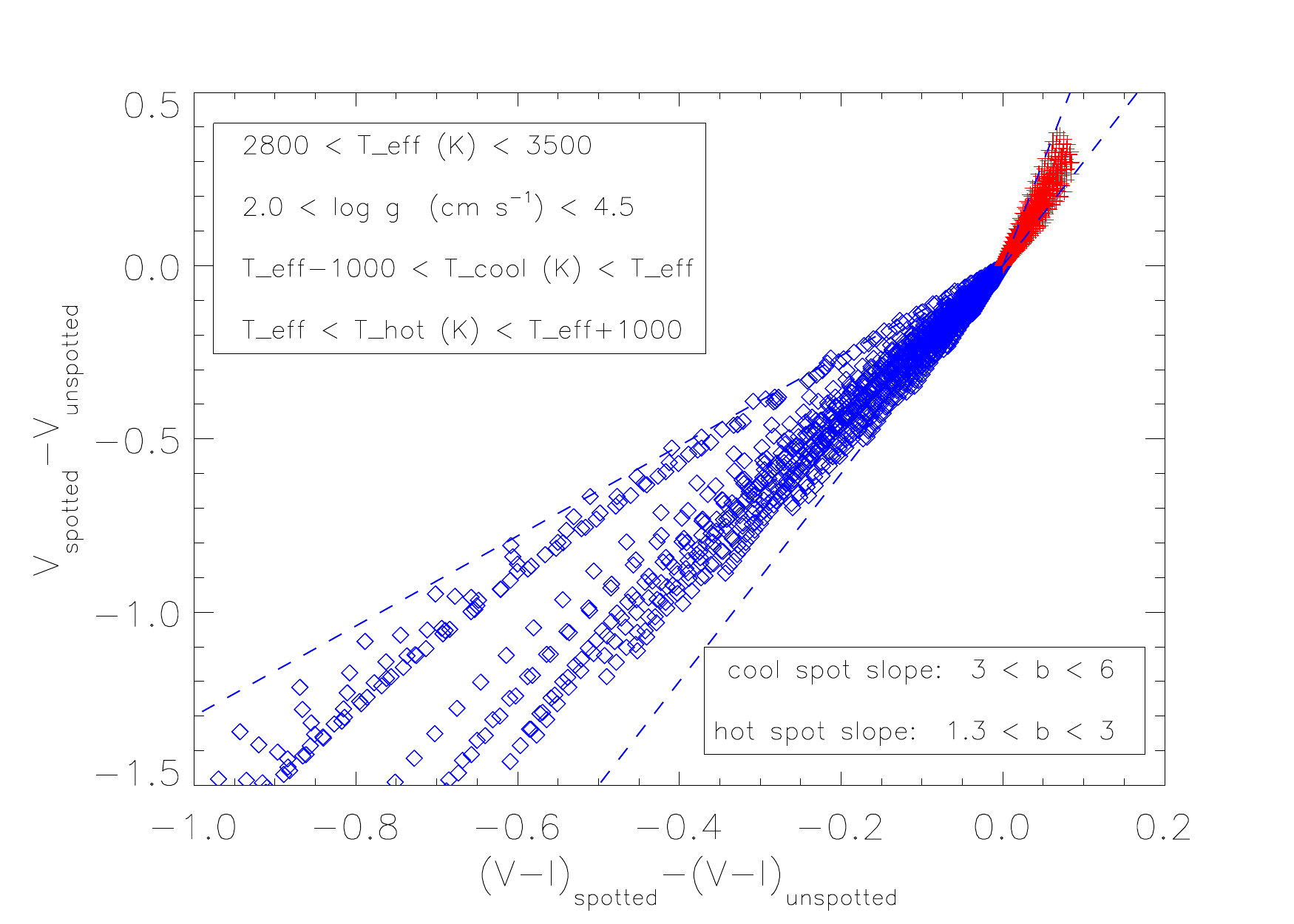,width=8cm,height=7cm,angle=0,trim = 80 0 0 0}
}
\end{minipage}
\caption{\label{model} Model magnitude versus color variations for a range of values of effective temperature, temperature contrast between perturbed and unperturbed photosphere and surface gravity.}
\end{figure}

\subsection{Correlation analysis}

Magnitudes  and colors  of ONC  stars  are affected  by both  the presence of magnetic
activity and  star-disk interaction effects. In order  to position our
stars  in  color-magnitude diagrams  we  should  use both  unperturbed
magnitudes  and   colors  (or   values  properly  corrected   for  the
activity/accretion effects at least).  If only cool spots were present
on the stellar surface, then  the brightest magnitude and bluest color
could  be   presumably considered intrinsic values linked with immaculate star.  This circumstance
may  possibly  apply to  most  WTTS  that  are  expected to  lack
accretion hot  spots. However, also in  WTTS there may  be presence of
hot faculae of  magnetic origin which can make  the star bluer, making
the   brightest    values   not   corresponding    to   an   unperturbed
level. Moreover, in the case  of binary systems with two components of
different  effective temperatures,  when the  more  active  and later-type  star
becomes fainter, owing to its variable activity, then the whole system
becomes bluer  (see, e.g. \citealt{Messina08}).   On the other  hand, if only
hot  spots were  present on  the  stellar surface,  then the  faintest
magnitude  and reddest color  could be  assumed as  'unspotted' values
(see, \citealt{Messina16}).  Both regression and correlation analyses
between magnitude  and colors  can allow us  to better  understand the
patterns of  color variation of  ONC stars and better  investigate the
nature of surface inhomogeneities.

\subsection{Model}

We  use  the  approach  proposed  by \citet{Dorren87}  to  simulate  the
amplitude of the V, R, I  magnitudes, and V$-$R, R$-$I and V$-$I color
variations  arising from  the  difference of  fluxes between  opposite
stellar  hemispheres  owing to  the  presence  of surface  temperature
inhomogeneities.    We note that our model assumes a two temperature photosphere, however, this is a simplified approach, the photosphere of T Tauri stars being likely more complex. \rm
The stellar  fluxes  were determined  by using  the
NextGen  atmosphere models  of  \citet{Hauschildt99} for  solar
metallicity  and convolved  with  the passbands  of  the Bessel  UBVRI
system (\citealt{Bessell90}).   Limb-darkening coefficients, different for the
unperturbed and the perturbed  photosphere, are taken from \citet{Diaz-Cordoves95}.   We  have  computed the  model  magnitude and  color
variations  for a  grid of  values  of temperature  and  covering
fraction assuming a two temperature photosphere.  More specifically, the hotter zone might be an accretion spot or faculae zone or it might just be the photospheric temperature. Similarly, the cool zone might be a region of magnetically inhibited convection or it might just be the photospheric temperature. \rm
The photospheric stellar magnitude and colors
are computed  over the effective  temperature range of the majority of our targets \rm (2800 $<$  T$_{\rm eff}$
(K) $<$  3500) and for a range  of surface gravities (2.0  $<$ log $g$
$<$  4.5).   The covering fraction was varied from
0.0 (no temperature inhomogeneities) to 1.0 (stars fully covered by temperature inhomogeneities)  \rm with a  0.1 increment,  whereas the  temperature  of the
surface inhomogeneities was  varied in the range T$_{\rm eff}$-1000  $<$  T
(K) $<$  T$_{\rm eff}$+1000   (see, e.g., \citealt{Berdyugina05}) \rm with a
100\,K increment.   In  our modeling  approach, gravity-darkening  effects are
neglected by  considering that these  effects tend to cancel  out when
computing the  flux difference between opposite  hemispheres.  We have
plotted in Fig.\,\ref{model} the results of our modeling. We note that
the amplitude of both magnitude  and color variations mostly depend on
the filling  factor of surface inhomogeneities, whereas  slopes of the
magnitude  vs. color  variation  primarily depend  on the  temperature
contrast between  such inhomogeneities and  the unspotted photosphere.
The presence of regions either cooler or hotter determines magnitude and color
variations  whose slope  $\Delta$\,mag/$\Delta$\,color is  positive in
both cases: cooler regions make the star fainter and redder (see the solid
red lines marked by crosses  in top right parts of Fig.\,\ref{model}),
whereas hotter regions make the star brighter and bluer (both magnitude and
color variations are  negative as shown by solid  blue lines marked by
diamonds in the bottom left  parts of Fig.\,\ref{model}). We find that
the  families of slope  obtained by  varying both  the star's  and the
perturbed region's parameters  fall within  a limited range  of values.   The most
interesting result  of our  modeling is that  slopes arising  from the
presence of only  regions cooler   (3 $<$  b $<$ 5) do not  overlap with the
slopes (1.3 $<$ b $<$ 3)  arising from the presence of only regions hotter than the unperturbed photosphere.
This allows  us to use the  V vs V$-$I color  variation to investigate
the nature of  temperature inhomogeneities supposed to be present in the stellar surface. In our
simulations  we   find  that  combinations  of  hot   and  cool  regions
simultaneously present on the same stellar hemisphere gives rise to slope
values close to zero or negative.

\begin{figure}
\includegraphics[width=95mm,height=95mm,angle=0]{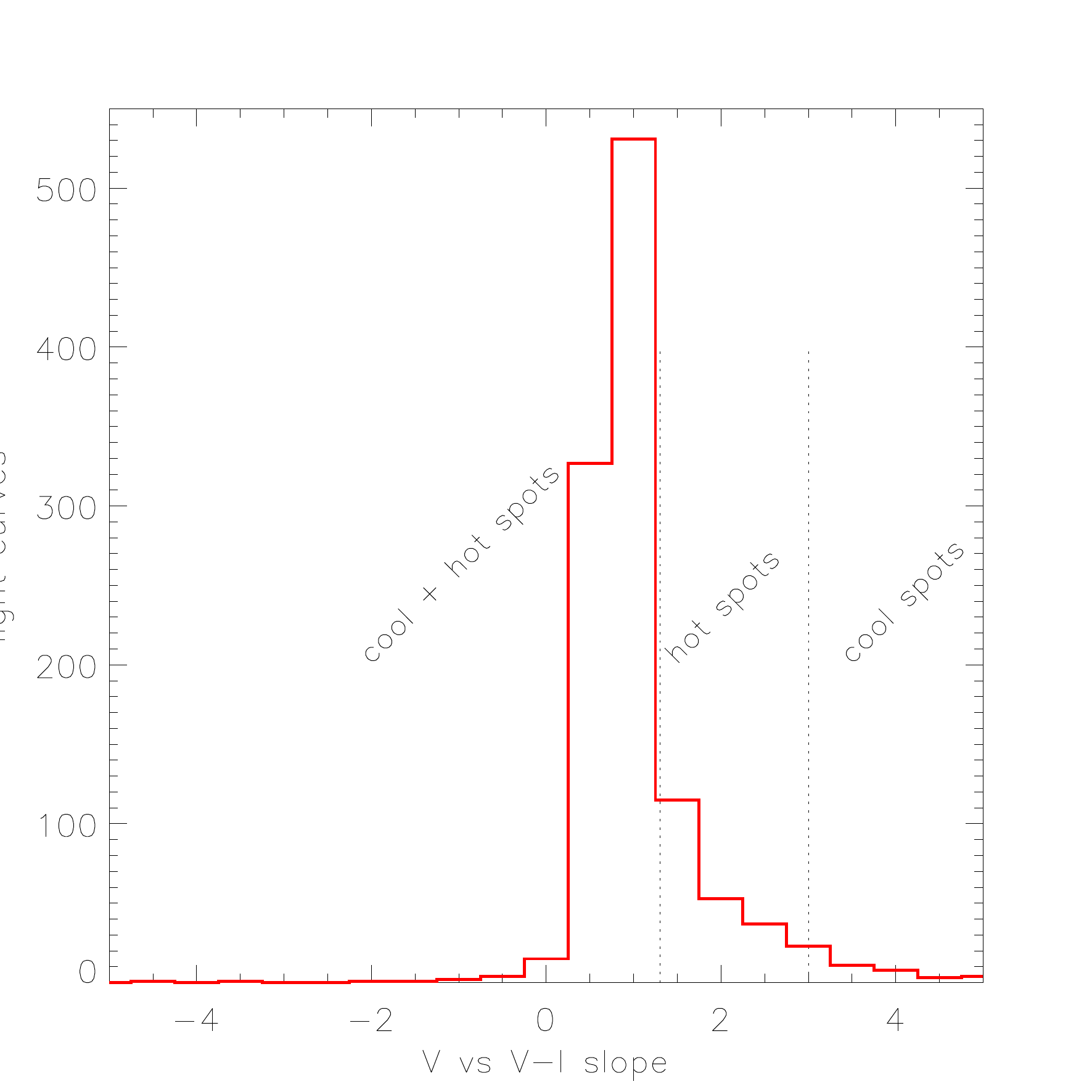}
\caption{Distribution of V versus V$-$I slopes for known WTTS and CTTS in our sample. The vertical dashed line demarcate  regions where according to our model, only cool spot (slope $>$ 3) only hot spot ( 1 $<$ slope $<$ 3) and a combination of cool and hot spots (slope $<$ 1) may exist.}
\label{histogram}
\end{figure}   

To  compare the  model  results with  observations,  we have  computed
linear  regression, correlation  analysis, and  computed  the relation
slopes,  their  correlation   coefficients,  and  the  Pearson  linear
significance levels associated to the correlation coefficients for the
V vs. V$-$I,  R vs. R$-$I, and V  vs. V$-$R magnitude-color relations.
The  results are  listed  in Table\,\ref{correl},  where  we list  the
following   information:    internal   identification   number   (id),
observation  season (seas),  mean  HJD (HJD$_{\rm  mean}$), number  of
measurements in the I, R and V band (N$_{\rm I}$, N$_{\rm R}$, N$_{\rm
V}$) slope of the V vs. V$-$I relation (b$_{\rm VVI}$) and correlation
coefficient  (r$_{\rm  VVI}$),  slope  of  the R  vs.  R$-$I  relation
(b$_{\rm RRI}$) and correlation coefficient (r$_{\rm RRI}$), and slope
of  the   V  vs.  V$-$R  relation  (b$_{\rm   VVR}$)  and  correlation
coefficient (r$_{\rm VVR}$), and slope  of the R$-$I vs V$-$R (b$_{\rm
RIVR}$) and  correlation coefficient  (r$_{\rm RIVR}$). We  listed and
used  in  the following  analysis  only  those  values for  which  the
correlation has a confidence level $>$ 95\%.\\

In  Fig.\,\ref{histogram} we  plot  the distribution  of  V vs.  V$-$I
slopes  of all light  curves. For  a number  of stars  we have  at our
disposal up to 5 values of  slopes determined from data collected in 5
observation seasons.  We verified that the slope distribution obtained
by considering all  the light curves and the  distribution obtained by
using for  each star  the mean slope  do not differ  significantly. We
also verified  that no  difference exists when  we consider  the whole
sample of stars  and a sub-sample of periodic  stars. Finally we opted
to consider the whole sample of stars (periodic plus non periodic) and
the whole sample of light curve in order to have a better statistics.

We find that about 6\% of  light curves show a slope consistent with a
variability arising from  only cool spot (i.e. regions of magnetically inhibited convection); 18\% of  light curves show a
slope  with a  variability arising  from  only hot  spot (i.e. either accretion spot or faculae zones); whereas  the
majority (76\%) show slopes consistent with a variability arising from
the presence of both cool and hot spots. We remind that periodic clean
and  periodic dirty stars  have a  mean slope  in the  range 1-5--2.2,
whereas  non periodic  stars,  which are  the  majority, have  smaller
slopes.  Moreover,  bluer stars  tend to have  slopes more  steep than
redder   stars.\\  This   distribution  provide   us   an  interesting
interpretation  about the  cause of  variability, that  is  stars with
either  only cool  or  only hot  spots  have inhomogeneities  patterns
sufficiently stable  to produce  periodic light variations.   One such
case is represented by the ONC  member V1481 Ori (\citealt{Messina16})
whose photometric variability is dominated by hot spots, but its light
curves were found  to be so smooth that we  could measure its rotation
period in 15 out 16 seasons. On other hand when the variability arises
from  the  combined effect  of  hot  and  cool spots  then  rotational
modulation gives very  scattered light curves and the  stars are found
to be non periodic.

\begin{figure*}
\includegraphics[width=55mm,height=85mm, angle=90]{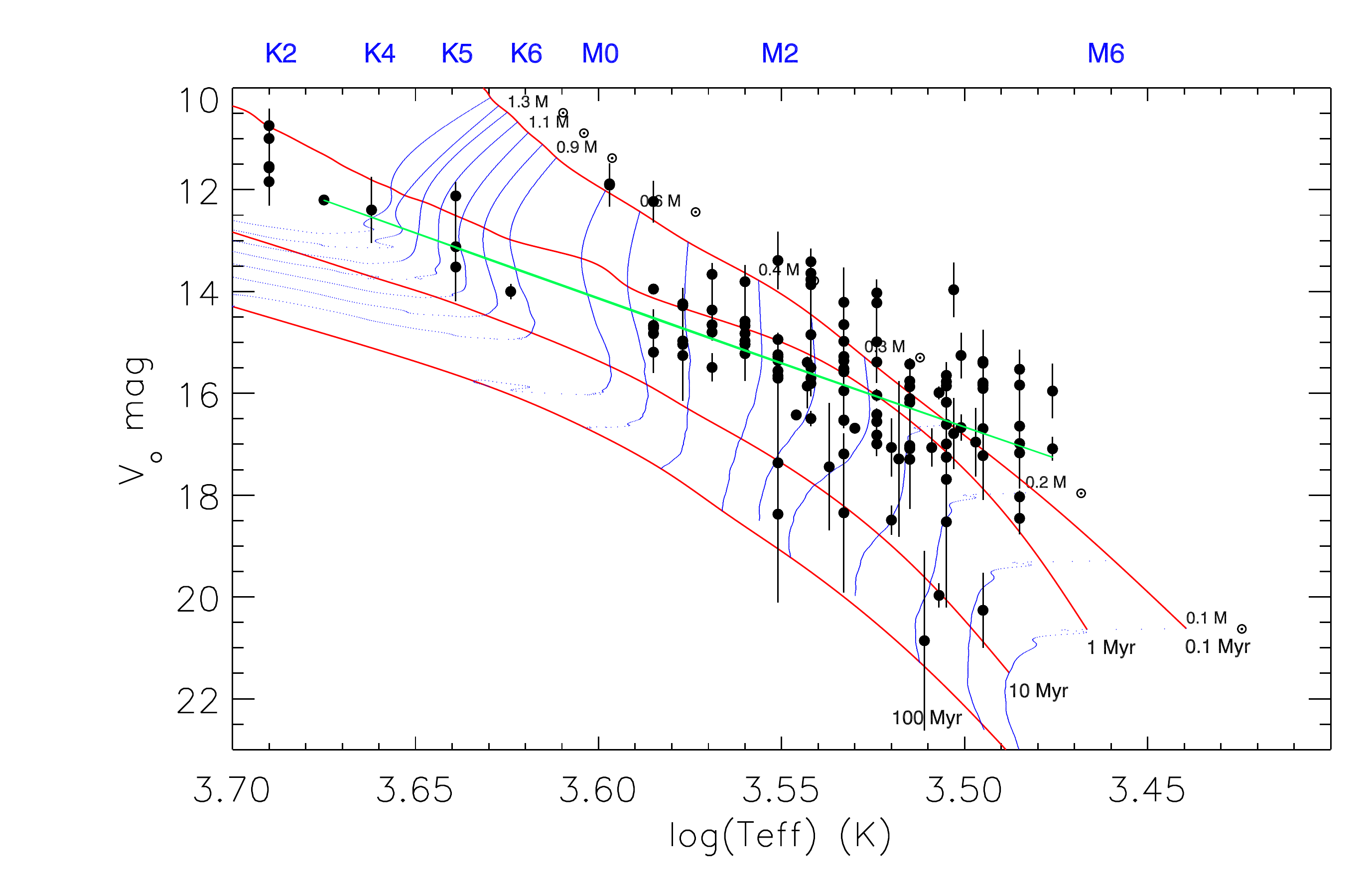}
\includegraphics[width=55mm,height=85mm, angle=90]{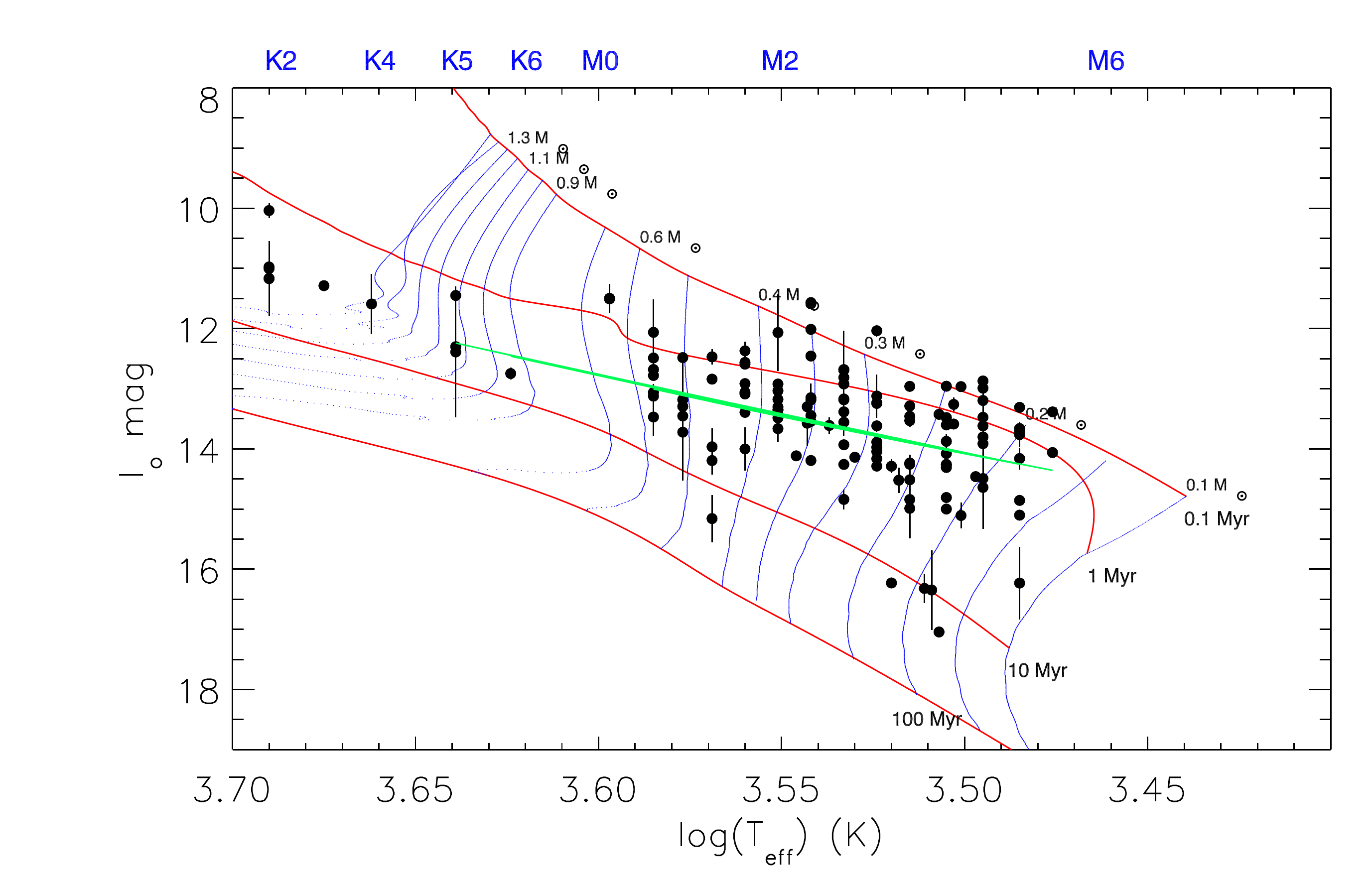}
\vspace{1cm}
\caption{T$_{\rm eff}$-V$_0$ (left panel) and T$_{\rm eff}$-V$_0$ (right panel) diagrams of the ONC members under analysis for which the extinction is known. The black bullets represent the mean magnitudes (average magnitudes over a 5-yr long time series) whereas the vertical bars their range of variation (brightest minus faintest magnitudes). Red solid lines and blue dotted lines represents theoretical isochrones and mass tracks, respectively, from \citet{Siess00}. }
\label{hr_vi}
\end{figure*}

\begin{figure*}
\includegraphics[width=55mm,height=85mm, angle=90]{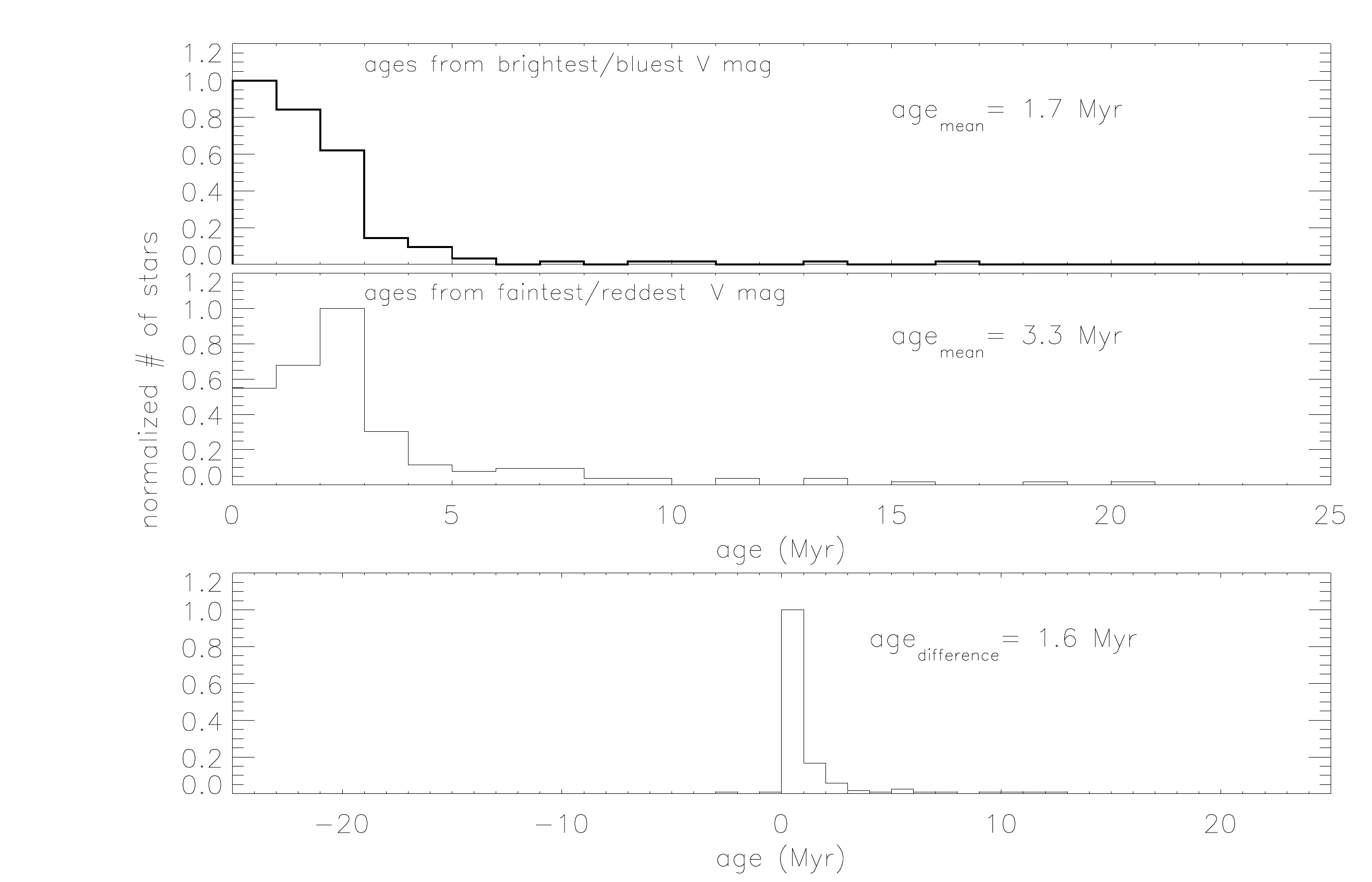}
\includegraphics[width=55mm,height=85mm, angle=90]{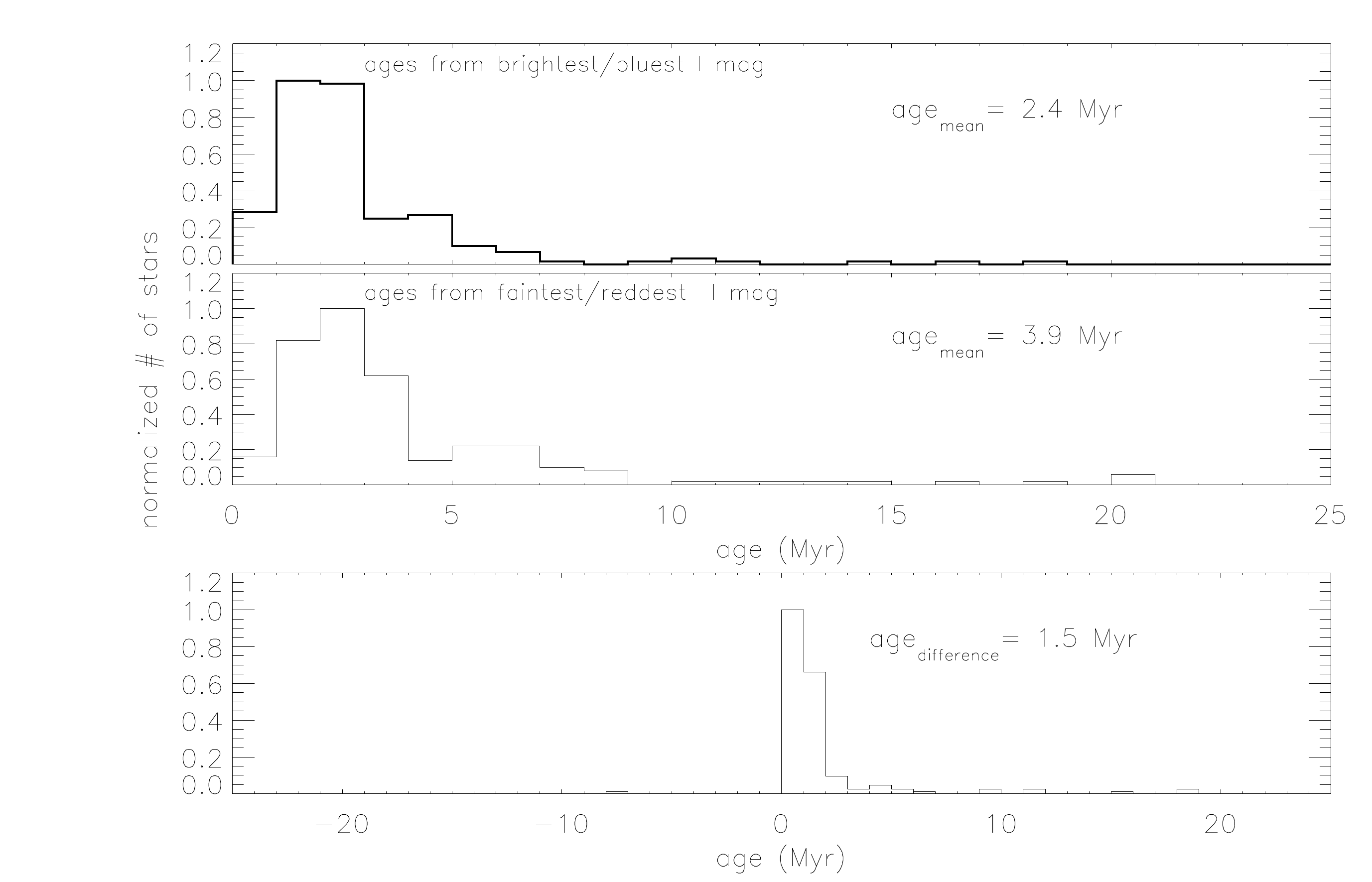}
\vspace{1cm}
\caption{Comparison between the mean magnitude dispersion in the T$_{\rm eff}$-V$_0$ diagram and the amplitude of photometric variability (V band on left panel and I band on right panel).
The magnitude dispersion is measured considering the residuals with respect to a 2nd order fit to the V$_0$ vs. T$_{\rm eff}$ distribution.  The amplitude of photometric variability is measured considering the brightest minus faintest magnitudes over a time range of 5 yr.}
\label{age}
\end{figure*}

\begin{figure*}
\includegraphics[width=55mm,height=85mm, angle=90]{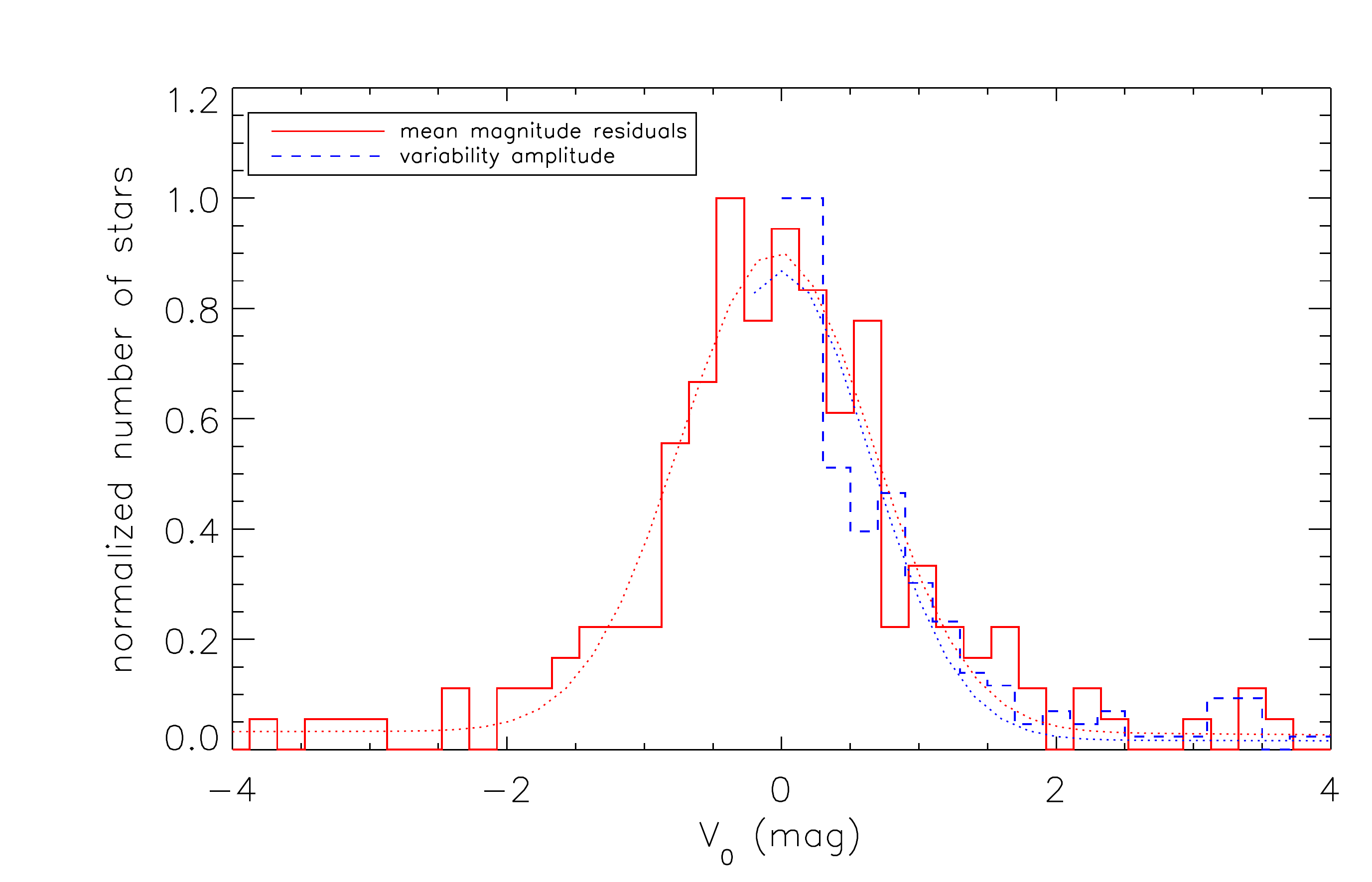}
\includegraphics[width=55mm,height=85mm, angle=90]{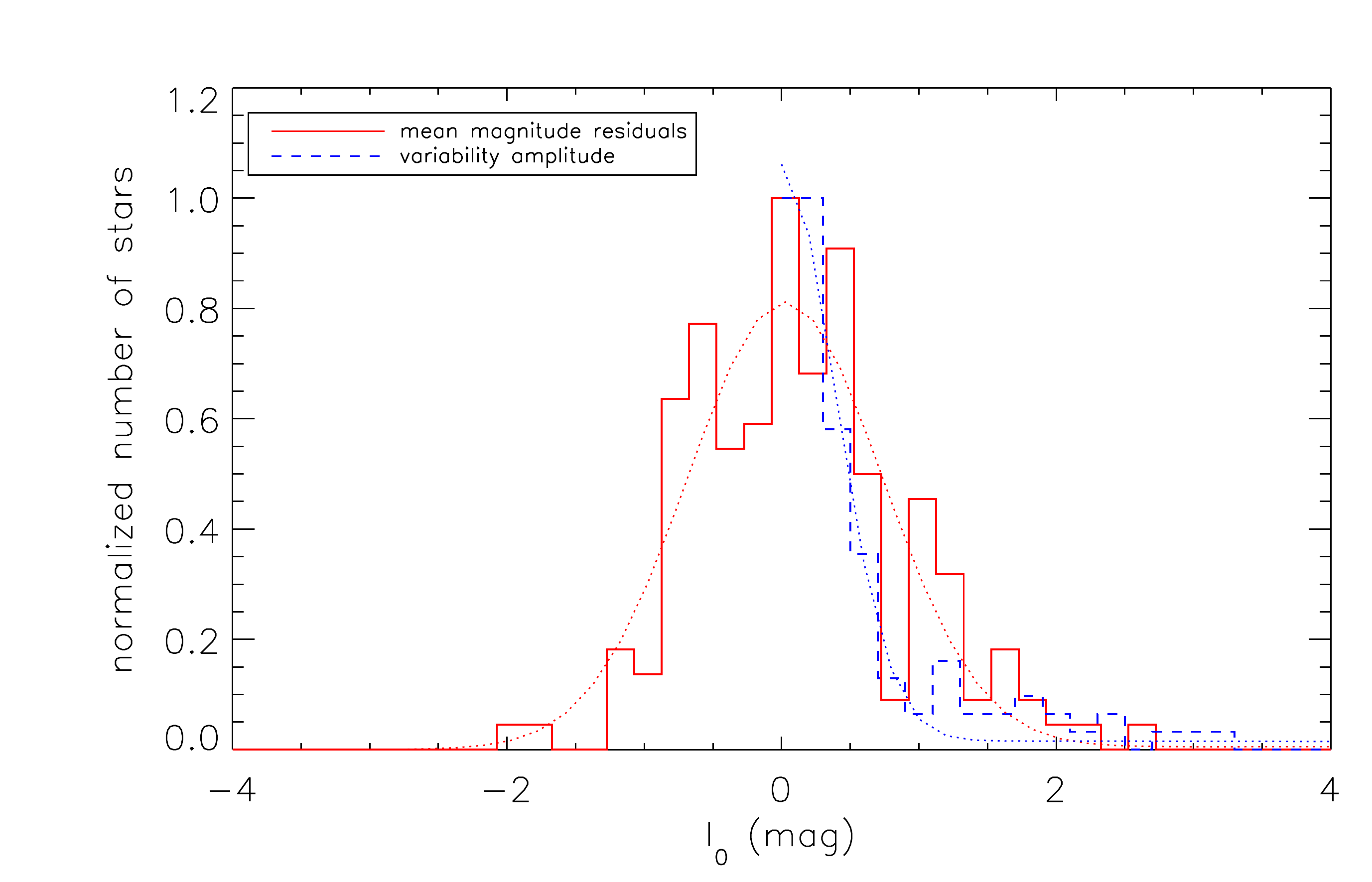}
\vspace{1cm}
\caption{Comparison between the mean magnitude dispersion in the T$_{\rm eff}$-V$_0$ diagram and the amplitude of photometric variability (V band on left panel and I band on right panel.
The magnitude dispersion is measured considering the residuals with respect to a 2nd order fit to the V$_0$ vs. T$_{\rm eff}$ distribution.  The amplitude of photometric variability is measured considering the brightest minus faintest magnitudes over a time rage of 5 yr.}
\label{hr_i}
\end{figure*}

\section{Age versus magnitude dispersion}

As mentioned earlier,  the prime  motivation  of present  study is  not to determine  absolute age   of the ONC  members under  analysis, but to explore  the  impact
of   photometric variability   on  inferred  stellar age and mass. Because of 
this, only one set of PMS evolutionary models well accomplished the aim.\\

We  found that  ages inferred  by means  of CMDs  are not  affected by
photometric  variability  if  stars  exhibit  a  periodic  photometric
variability and the rotation  phase magnitude dispersion is negligible
with respect to  the peak-to-peak light curve amplitude  ($R \ge 2.5$,
the  so  called  'clean  periodic').  CMDs still  provide  a  reliable
estimate of the  age also for periodic variables  whose rotation phase
magnitude  dispersion  becomes comparable  to  the peak-to-peak of  light
curve amplitude (1 $\le R < 2.5$, the so called 'dirty periodic'). The
inferred  ages in the second   case is found to be slightly older  and the  number of
outliers  (stars  with  ages  significantly older  than  the  average)
reaches the 20\%. However, the majority of stars in our sample are non
periodic and ages inferred from CMDs become unreliable.\\

Very accurate  knowledge of   the photometric variability
and  the use  of  the magnitude-T$_{\rm  eff}$  diagrams (to  overcome
uncertainties  arising  from  the   color  variability)  can be a robust tool 
to check the  reality of  age spread among the ONC
members which has been debated by several  authors (see, e.g., \citealt{Reggiani11}).  In
the  following, we  estimate  the age  of  the ONC  members using  the
magnitude-T$_{\rm eff}$  diagrams, where the color is  replaced by the
spectroscopically  measured  effective  temperature.  We  focus  on  a
sub-sample of  stars with known reddening that  comprises periodic and
non   periodic   members.   In   the   left   and   right  panels   of
Fig.\,\ref{hr_vi},  we  plot  the  mean  V$_0$  and  I$_0$  magnitudes
(averaged over  the 5-yr time base) versus  the effective temperatures
taken from \citet{Hillenbrand97}. The vertical bars indicate the observed
maximum  range  of  magnitude  variation.  Solid red  and  blue  lines
represent isochrones for  the ages from 0.1 Myr to  100 Myr taken from
\citet{Siess00} and the evolutionary mass tracks in the range from
0.1\,M$_\odot$  to 1.3\,M$_\odot$.   In Fig.\,\ref{age},  we  plot the
distributions  of  ages   obtained  considering  the  brightest/bluest
magnitudes,  faintest/reddest magnitudes,  and the  difference between
the oldest and youngest ages for  each star.  We find that the average
ONC  age can be  1.7\,Myr (brightest/bluest)   to 3.3\,Myr (faintest/reddest)  and that  the photometric variability can produce an average age difference of 2.5\,Myr in the V
band, and slightly larger in the I band.\\

In  order to  quantify the  age dispersion,  we measure  the magnitude
residuals  with  respect to  a  second  order  polynomial fit  in  the
magnitude--T$_{\rm eff}$ diagram. The distribution of the residuals is
plotted  in Fig.\,\ref{hr_i} and  represented with  a heavy  solid red
line.   We compare  this  distribution with  the  distribution of  the
amplitudes of photometric variability overplotted with an heavy dashed
blue line. A  gaussian fit to the distributions  (dotted red and blues
lines,  respectively) is used  to measure  the standard  deviations of
both  distributions.   We  find  that the  distribution  of  magnitude
residuals with  respect to the  polynomial fit has  standard deviation
$\sigma$ =  0.70\,mag, which is  comparable to the  standard deviation
$\sigma$ =  0.64\,mag of  the distribution of  variability amplitudes.
This new results  derived from 5-yr V band  photometry contradict with
the  conclusions made by  \citet{Reggiani11}  that observational
uncertainties   and   variability   cause  a   luminosity   dispersion
significantly smaller  than the apparent age dispersion.  We find that
that V band photometric variability  level is large enough to mask any
age spread and may be sole cause of the observed magnitude spread.

In the  case of  the I  magnitude, we find  that the  distributions of
magnitude  residuals  and   amplitudes  have,  respectively,  standard
deviations  $\sigma$  = 0.71\,mag  and  $\sigma$  = 0.39\,mag.   Which
indicates that  the observed  photometric variability can  account for
only  a part  ($\sim$50\%) of  the  observed magnitude  spread in  the
magnitude-T$_{\rm  eff}$ diagrams.  To account  to the  remaining 50\%
spread,  an intrinsic age  distribution among  the ONC  member becomes
essentials.  The  other possibility can  be a radius  dispersion among
coeval  members  as  consequence  of non-uniform  accretion  at  early
stages, as suggested by \citet{Jeffries11}.

 As  mentioned  several  times,  our  statistics  on  the  variability
provides  only a lower  limit owing  to the  limited extension  of the
observation  data  base  and  the  insufficient  time  resolution  for
transient phenomena. However,  as shown in Fig\,\ref{rms_boundary}, we
do not  expect that  the variability amplitudes which  we have  measured are
significantly  underestimated.  The inferred  age  dispersion, if  not
real, must be imputed to some other effects beside the variability.

\section{Conclusions}

We have monitored a sample of 346  members of the ONC in the V, R, and
I bands  for 5 consecutive  years during 2003-2008.  From  analysis of
the  multi-band  photometric  time   series  data,  we  could  measure
variability in  magnitude and  color for all  stars, which  shows wide
range of amplitudes.  We measured the magnitude and colors maximum and
minimum values, and  the amplitude of variation, of  each star in each
observation season and in the  full 5-yr timeseries.  We find that the
amplitude of  variability in  the V  band is on  average a  factor two
larger than in the I band. Moreover, the measured level of variability
is   found   to   increase   as   we   extend   the   base   line   of
observation.  However, we find  that within  2-3 consecutive  years of
monitoring,  we could capture  almost full  range of  the variability.
The magnitude and color maximum  and minimum values were used to infer
mass and age of each star in color-magnitude diagrams using isochrones
and   evolutionary  mass   tracks  from   the  models   of   \citet{Siess00}. We  found that  stars  showing periodic  and very  smooth
rotational  light variations  undergo magnitude  and  color variations
with same slope  as the theoretical isochrones. As  a consequence, the
variability does  not affect  their inferred mass  and age.   On other
hand,  stars  showing periodic  but  very  scattered rotational  light
variations  or   non  periodic  stars  undergo   magnitude  and  color
variations with  a slope smaller  than that of  theoretical isochrones
and, consequently,  appear older and more massive.  Modeling the slope
of  the  magnitude  versus  color  variations  by  following  the  the
methodology given  by \citet{Dorren87}, we  find that the  use of  V vs.
V$-$I  variations allows to  infer if  the photometric  variability is
caused by zones hotter, zones cooler than the unperturbed photosphere, or a combination of both.  We find the
majority   of   stars   to   fall   in  the   last   category.   Using
magnitude-T$_{eff}$ diagrams, instead of color-magnitude diagrams where color
is affected  by significant variability,  and the PMS  models of
\citet{Siess00}, we find  that for V band the apparent age spread can
be mostly accounted by  the photometric variability. In other words
the  age dispersion  in the  ONC is  completely masked  by  the V-band
variability. On the  contrary, the amplitudes of variability  in the I
band have  a dispersion that is a  factor two smaller than  the I mean
magnitude dispersion.  Therefore,  I-band variability can take account
of only  50\% of  magnitude spread in  the HR diagram.   The remaining
50\% spread  requires a  different explanation and  may be due  to age
dispersion within ONC members which we find to be of about 1.5\,Myr.

\section*{Acknowledgements}
Research on stellar activity at INAF- Catania Astrophysical Observatory is supported by MIUR  (Ministero dell'Istruzione, dell'Universit\'a e della Ricerca). The work is based on data collected at the Indian Institute of Astrophysics (IIA). The extensive use of the SIMBAD and ADS databases operated by the CDS centre, Strasbourg, France, is gratefully acknowledged. We thank the referee, Prof. W. Herbst, for valuable comments.\\

\bibliographystyle{mnras} 
\bibliography{biblio}

\appendix 

\section{Measurement of light curve amplitude}
\label{AppendixA}

\begin{figure}
\includegraphics[width=90mm,height=140mm]{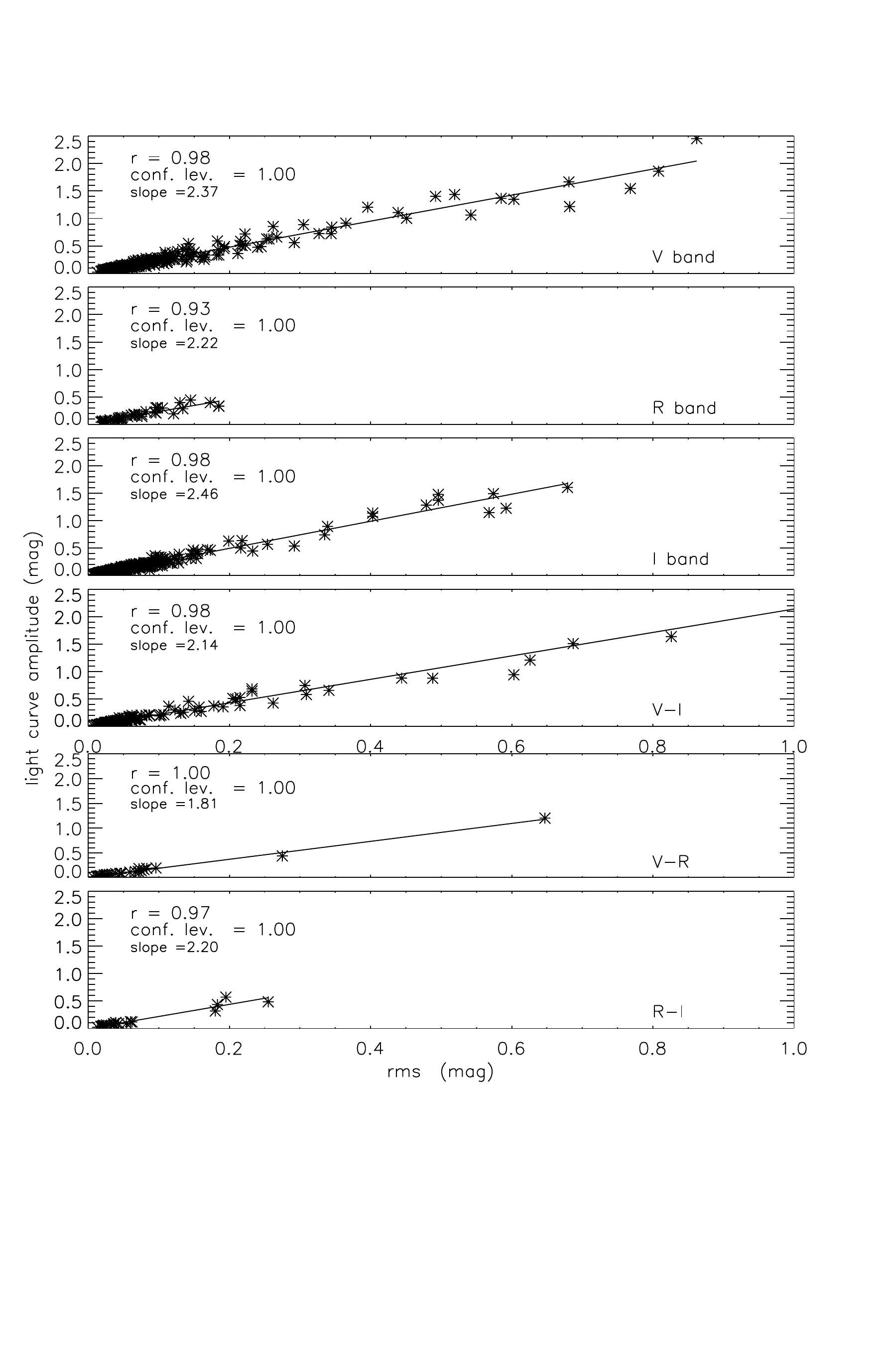}
\vspace{-3cm}
\caption{Distributions of light and color curve amplitudes versus RMS. Solid lines are the linear fits. Labels report the correlation coefficient(r), the confidence level from Pearson test, and slope of the linear fits.}
\label{corr_ampl}

\end{figure}

As described earlier,  to measure the amplitude of  the clean periodic
light curves we use the amplitude  of the sinusoidal fit. This ensures that
there   is   no   overestimation   of  amplitude   due   to   residual
outliers. Whereas, in  case of dirty periodic stars,  the amplitude of
the sinusoidal  fit tends to  underestimate the true amplitude  of the
light curve.  Therefore, for the dirty periodic  and irregular variables
we  estimate the  brightest and  faintest  magnitudes as  well as  the
amplitude  by  following  two  different  approaches.   In  the  first
approach we derive the relation  between the light curve amplitude and
its root mean square (RMS)  using the sample of clean stars.  Thereafter,
we apply this relation to estimate the amplitude of the light curve of
dirty stars  by using the  computed rms.  In  Fig.\,\ref{corr_ampl} we
have shown the  plots of amplitude versus RMS  for different bands and
colors, whereas  fitting results are given in  the Table \ref{ratio}.
In each panel we  report the correlation coefficient, its significance
level according  to the  Person statistical test,  and the slope  of a
linear fit.  We found very tight correlation between amplitude and RMS
and  this relation enables  us to  derive the  average light  and color
curve amplitudes from the measurement of their RMS for the stars which
have either  no smooth  light curves or  having light curves  but very
scattered  ('dirty'  periodic  sample).   To minimize  the  effect  of
outliers  on the  determination of  RMS, we  select only  light curves
which have at least 30 measurements.\\

Another approach  is to  consider the, e.g.,  20\% brightest  and 20\%
faintest magnitudes of a 'dirty'  light curve and to use their average
values as  measurement of the brightest and  faintest magnitude.  Both
approaches can be  tested with clean light curves  which has got known
amplitudes  of  variability.  That  means  we  can compare  amplitudes
derived from direct sinusoidal fitting to amplitudes computed from RMS
(first  approach)  and/or the  amplitudes  obtained  by averaging  the
extreme bands of the magnitude distribution (second approach).\\

We  find  that  the   amplitudes  derived  from  both  approaches  are
essentially  similar, i.e,  their difference gives  zero mean
value with a dispersion of about 0.01\,mag.  As expected, we find that
increase in  magnitude of  phase dispersion of  the light  curve (from
clean to dirty) deliberately makes the amplitude of the sinusoidal fit
to be smaller than the amplitude  derived from either the first or the
second approach.\\

We have also made a comparison  between the distribution of RMS in the
clean  and   dirty  samples   to  further  investigate   whether  they
statistically  belong  to   same  or  to different  distributions.   The
confidence  level that  RMS  distributions of  light  curves in  the
'dirty' and 'clean' samples come  from the same parent distribution is
computed with Kolmogorov-Smirnov (KS) tests. We find that it is larger
than  99.9\%  in  the I  band,  53\%  in  the  R  and 13\%  in  the  V
band. Although the R band distribution is based on only 41 against the
243 values of V band, its  confidence level is much higher. In case of
V  and R bands  we find  that distributions  are quite  different which
appears not due to the  poor statistics (smaller sample), but seems to
be  an effect   of  variability  phenomena  not   related  to  rotational
modulation.  A  comparison between  color  RMS distributions  reflects
similar results:  R$-$I color has RMS  in the clean  and dirty samples
more  similar  than  V$-$R  and V$-$I  (KS$_{\rm  ri}$=83\%,  KS$_{\rm
vi}$=35\%, KS$_{\rm vr}$=21\%).    Based on the results  of KS tests,
we transform  dirty RMS into amplitude  values only for I  and R bands
and R$-$I color. The V band amplitudes derived from RMS may be little
over estimated,  however that gives  us a possibility to  investigate the
impact  of magnitude and  color variations  in CMDs  on a  much larger
stellar sample.   We could verify that  no dependence of  the ratio on
the star's  brightness exists, since the  photometric accuracy equally
affect  both  the  light  curve  amplitude  and  the  dispersion.   To
summarize,  amplitudes of periodic  clean light  curve can  be derived
from sinusoid fit, RMS and  percentiles. Amplitudes of non periodic or
dirty  light  curves  are  better  estimated  from  percentile,  since
sinusoid fit  underestimated whereas RMS method  overestimate.  To get
homogeneous estimated  we have used  the percentile for all  the light
curves.

\begin{table}
\caption{Average ratio between the  amplitude and RMS of light and color curves..\label{ratio}}
\begin{tabular}{l c c}
Band & amplitude/rms & \# light curve\\
\hline
V & 2.37 $\pm$0.03 & 235 \\
R & 2.22 $\pm$0.14 & 41 \\
V & 2.46 $\pm$0.02 & 350\\
V$-$I & 2.14 $\pm$ 0.03 & 350 \\
V$-$R & 1.81 $\pm$ 0.03 & 27\\
R$-$I  & 2.30 $\pm$ 0.12 & 25\\
\hline
\end{tabular}
\end{table}

\end{document}